\documentclass[aps,prl,preprint,superscriptaddress]{revtex4-1} 
\usepackage{framed,multirow}
\usepackage{braket}
\usepackage{amsmath}
\usepackage{amssymb}
\usepackage{latexsym}
\usepackage{url}
\usepackage{xcolor}
\usepackage{graphicx}

\selectcolormodel{gray}
\graphicspath{{img-gray/}}
  
\begin{document}

\title{Determining electronic properties from $L$-edge X-ray absorption spectra of transition metal compounds with artificial neural networks}

\author{Johann L{\"u}der}
\email[]{johann.lueder@mail.nsysu.edu.tw}
\affiliation{Department of Materials and Optoelectronic Science, National Sun Yat-sen University, 70 Lien-Hai Rd., Kaohsiung 80424, Taiwan, R.O.C.}
\affiliation{Center for Crystal Research, National Sun Yat-sen University, 70 Lien-Hai Rd., Kaohsiung 80424, Taiwan, R.O.C}
\date{\today}

\begin{abstract}
X-ray absorption spectroscopy at the $L$-edge probes transitions of 2p-electrons into unoccupied d-states. Applied to transition metal atoms,  this experimental technique can provide valuable information about the electronic structure of $d$-states. However, multiplet effects, spin-orbit coupling, a large number of possible transitions can cause a rather involved nature of  2$p$ XAS spectra, which can often complicate extracting of information directly from them.  Here, artificial neural networks trained on  simulated spectra of a 2$p$ XAS model Hamiltonian are presented that can directly determine information about atomic properties and the electronic configuration of d-states from $L$-edge X-ray absorption spectra. 
Moreover, the adaptable nature of artificial neural networks (ANNs) allows extending their capability to obtain information about the electronic ground state and core hole lifetimes from 2$p$ XAS spectra as well as to incorporate external factors, such as temperature and experimental convolution that can affect details in spectral features.
The  effects of noise and background contributions in spectra on the accuracy of ANNs are discussed and the method is validated on experimental spectra of transition metal compounds, including metal-organic molecules and metal oxides. 

\end{abstract}


\maketitle

\section{Introduction}

X-ray spectroscopy (XS) is an important tool for many branches of science and technology \cite{Ponrouch2016,Hansmann2008,Wackerlin2010}.
Since the discovery of X-rays, a large variety of specialized XS techniques have been developed, and many of them are routinely used as scientific tools, providing detailed information about properties of materials that lead to accelerated technological and scientific advancements \cite{Hartmann2012,Stadtmuller2014,Verdini2012,Wakonig2019,Brumboiu2019a,Ma2019,Golze2018,Haverkort2010a}.  
  Many technological applications, such as environmental sensors, catalysts, and energy storage and harvesting devices, can benefit from these insights, which allow to understand and optimize processes in materials, e.g., involving details of nanostructures \cite{Daniel2004}, charge transfer effects \cite{Ghiasi2019a}, and chemical stability \cite{Wang2017268,Lim2005}, that can be employ to enhance their design and functionalization. 
One specialized technique is soft X-ray absorption spectroscopy (XAS) at the L-edge that employs tunable photon energies probing energy-dependent transitions of electrons from occupied to unoccupied states obeying Dipole Selection Rules \cite{Dirac1927}.  
This technique is element, state and orientation-selective. In light transition metal (TM) atoms, this technique can probe electron transitions from occupied 2$p$ to unoccupied 3$d$-states  induced by the absorption of photons,  and help to reveal the nature of their 3d states.
 
For  extensive analysis, experimental  XAS results are frequently combined with theoretical simulations  \cite{Rehr2000,Wilson2005,Rauls2012,Odelius2006}.
  Then, experimental and computational insights can explain possible causes leading to changes in spectra. Detailed pictures can be drawn from the underlying mechanisms, explaining the role of the electronic and chemical structure and relate them to changes in properties of materials \cite{Wang2010b,LucianoTrigueroYiLuo1999,Brena2006,Brena2004a,Weiss1998,Janczak2001}.  
However,   the independent particle methods, that are often used in other branches of theoretical spectroscopy \cite{Luo1998,Triguero1998,Zhang2017}, cannot describe  
2$p$  
XAS processes in TM ions, due to strong electron-hole interactions, multiplet effects and  spin-orbit coupling (SOC) \cite{de2008core,Groot2005}.
These effects govern the electronic transitions  in 2$p$ XAS and strongly  affecting spectral shapes, including the non-trivial branching ratio between peaks corresponding to transitions from $p_{3/2}$ and $p_{1/2}$ semicore states. 
Several theoretical many-body methods have been developed  to model 2$p$ XAS spectra at the $L$-edge, such as first-principle methods employing, e.g., the Bethe-Salpeter equation \cite{Rehr2006,Shirley2004,Shirley2005,Vinson2011}  
and configuration interaction related methods \cite{Kumagai2009,Josefsson2012}.
In contrast to first-principle methods,  XAS model Hamiltonians are computationally efficient and can yield very good agreement with experiments \cite{DeGroot2008,Stavitski2010}, although their parameterization can be challenging and time-consuming.
Recently, model Hamiltonian approaches (MHA) were combined with DFT results through extracting local properties using maximally localized Wannier functions  \cite{Haverkort2012} or extracting the parameterization from the hybridization function, which is known from the dynamical mean-field theory \cite{Luder2017a}. Still, none of these methods provides a direct way to determine electronic properties from spectra, but rather require computing the electronic structure, followed by comparing an obtained theoretical XAS spectrum with the experimental spectrum. Additionally, the possibly large number of transitions  can lead to feature-rich spectra that are difficult to analyse. 
 To   directly  analyse spectra,  a function is needed that reverses the conventional computation of $L$-edge XAS spectra performed with, e.g. a MHA. 
Then,  the model Hamiltonian can be used to determine physical properties such as the electronic configuration of the ground state and, in principle, also the theoretical $L$-edge XAS spectrum with the results given by the inverse function. This function is, however, unknown.  

 Artificial neural networks (ANNs) have attracted much attention in  material science \cite{Bhadeshia1999,Butler2018,Ji2017} because they can fit any real-valued function with arbitrary accuracy \cite{Cybenko1989}, and allow unbiased learning of fundamental relations from data.  
Employing ANNs may lead to solutions to complex problems, as ANNs can  accelerate computational methods and accelerate materials discovery \cite{Xia2018,Chmiela2017,Balabin2009,Kamath2018,Raccuglia2016,Kumar2020}.  
 For instance, ANNs are frequently combined with Density Functional Theory (DFT) computations \cite{Brockherde2017,Schleder2019,Chandrasekaran2019}  to efficiently predict structures and properties   of   compounds,  which can be computationally expensive by DFT means \cite{Oliynyk2016,Hansen2015,Mills2017}. With regard to 2p XAS spectroscopy, one question at hand is:  Is it possible to construct useful ANNs that represent this inversion and can obtain information about the electronic structure from experimental 2$p$ XAS spectra  in the form of parameters used in, e.g. a MHA? 

Here,  an ANN   method is presented that obtains information on the  
TM ion's local electronic  structure  from experimental $L$-edge XAS spectra in the form of parameters used in a MHA. The MHA is the basis of the ANN method. THe ANN can obtain relative energy levels of the d-states in a TM ion and screening factors of Slater-Condon integrals. Furthermore, extensions of the method provide estimations of core-hole lifetimes and can consider external factors such as temperature and  experimental broadening, as well as predictions of expectation values of the MHA, including spin state and orbital occupations. 
Several ANN architectures were tested to demonstrate limitations, capabilities and accuracy dependencies. The method is applied to and tested with previously published experimental 2$p$ XAS spectra of well-studied systems of transition metal monoxides and transition metal-organic complexes. 

 \section{Methods}

 In the following, an  ANN approach is described that can extract information about the local electronic structure of transition metal ions such as relative d-level positions and d-state occupations  from 2$p$ XAS spectra with high accuracy. Then, the influence of external factors such as temperature and experimental convolution as well as expectation values prediction (EP) is included through  extensions of the method. The ANNs are constructed and trained on simulated 2$p$ XAS spectra, the so-called reference spectra, computed for light TM ions.  The employed 2$p$ XAS MHA is based on crystal-field multiplet theory (CMT) while charge transfer (CT) effects are neglected.  Since 2$p$ XAS spectra of light transition metal have large energy separations (i.e. spectroscopic fingerprint), the approach works with specialized ANNs for each element. 
 The trained ANNs can  be applied to experimental spectra to extract sets of parameters  (that are atomic and ligand field parameters in the language of CMT) describing characteristic electronic properties. Then, the determined parameters allow  recomputing any spectrum with the same theoretical framework as the reference spectra to estimate the accuracy of the method directly.
In the following, CMT and the ANN implementations  are briefly explained.  Then, the performance of different ANN types is compared and the influence of noise and background in spectra is discussed before the method is applied to experimental spectra.
  
\subsection{A 2$p$ XAS Model Hamiltonian} 
For TM complexes with partly filled 3d shell, a  CMT  based approach is used to build a database of computed $L$-edge XAS spectra for  
Ni$^{2+}$, Co$^{2+}$, Fe$^{2+}$ and Mn$^{2+}$ ions.  
  CMT considers a central metal-ion described by valence 3$d$-states and 2$p$ semi-core states. The TM-ion is placed in a crystal-field (CF) representing the chemical surrounding, leading to the lifting of the degeneracy of 3$d$-states. The strength of the CF is described by optical parameters, for instance, in  Ballhausen notation \cite{Carlin1963}.
   The $d$-level splitting is then given for D$_{4h}$ symmetry by
  \begin{equation}
   \begin{split}
 \epsilon_{d_{a_{1g}}}  =   6  D_q -  2 D_s - 6  D_t \\
 \epsilon_{d_{b_{1g}}} =    6  D_q + 2 D_s - D_t  \\
 \epsilon_{d_{b_{2g}}} =   -4  D_q + 2 D_s - D_t  \\
 \epsilon_{d_{e_{g}}} =     -4 D_q  - D_s + 4 D_t \\
 \end{split}
\label{eq:d4h}
\end{equation}
For O$_h$ symmetries, the $d$-level split reads as 
  \begin{equation}
   \begin{split}
 \epsilon_{d_{e_{g}}}= 6 \cdot D_q \\
 \epsilon_{d_{t_{2g}} } = -4 \cdot D_q. \\
 \end{split}
\label{eq:oh}
\end{equation}
Negative values of $D_q$ in Eq.~(\ref{eq:oh}) yield a C$_4$ symmetric CF \cite{Carlin1963}. 
  The $L$-edge XAS model Hamiltonian \cite{Haverkort2012}  is
 \begin{equation}
 \begin{split}
\hat{H}=\sum_{i,j} \epsilon_{d_{i,j}} \hat{d}_i^\dagger \hat{d}_j  
+ \sum_{i,j} \epsilon_p \hat{p}_i^\dagger \hat{p}_j
+ \zeta_d \sum_{i,j} \braket{d_i|\vec{l}\cdot \vec{s}| d_j} \hat{d}_i^\dagger \hat{d}_j   \\
+ \zeta_p \sum_{i,j} \braket{p_i|\vec{l}\cdot \vec{s}| p_j} \hat{p}_i^\dagger \hat{p}_j   
+ \sum_{i,j,k,l} U^{dd}_{ijkl} \hat{d}_i^\dagger \hat{d}_j^\dagger \hat{d}_l \hat{d}_k
+ \sum_{i,j,k,l} U^{dp}_{ijkl} \hat{d}_i^\dagger \hat{p}_j^\dagger \hat{p}_l \hat{d}_k.
 \end{split}
\label{eq:ModelHAM}
\end{equation}
 In Eq.~(\ref{eq:ModelHAM}), the non-relativistic single-electron $d$- and $p$-state energies  are given by $\epsilon_d$ and $\epsilon_p$, respectively. The corresponding $\hat{d}_j$ ($\hat{d}^\dagger_j$),  and  $\hat{p}_j$ ($\hat{p}^\dagger_j$) operators  annihilate (create) states.   Spin-orbit coupling constants are  $\zeta_d$ for $d$-states and $\zeta_p$ for $p$-states. They are linked by the vector product $\vec{l} \cdot \vec{s}$  with the angular-momentum operator ($\vec{l}$) and the spin operator ($\vec{s}$). 
  The onsite Coulomb interaction between $d$-electrons can be described by Slater-Condon integrals $F^0_{dd}$, $F^2_{dd}$ and $F^4_{dd}$. Likewise, the interaction between $d$-electrons and $p$-core hole  are described by $F^0_{pd}$, $F^2_{pd}$, $F^4_{pd}$, $G^1_{pd}$ and $G^3_{pd}$ \cite{Groot2005}. 
  The monopol terms between 3d-electrons, i.e. $F^0_{dd}$, and  between 3d and 2p-electrons, i.e. $F^0_{pd}$, are expressed through  $U^{dd}_{ijkl}$, $F^2_{dd}$ and $F^4_{dd}$, and $U^{dp}_{ijkl}$, $G^1_{pd}$ and $G^3_{pd}$, respectively \cite{haverkort2005spin}, where U's are Hubbard-U parameters; they read

\begin{equation}
  U_{dd}= F^0_{dd} - \frac{2}{ 63} ( F^2_{dd} +F^4_{dd})
  \label{eq:udd}
\end{equation}
and
  \begin{equation}
  U_{pd}= F^0_{pd} - \frac{1}{15}  G^1_{pd} - \frac{3}{70} G^3_{pd}.
  \label{eq:upd}
\end{equation}
    The values of the Slater-Condon integrals other than $F^0_{dd}$ ($F^0_{pd}$) are taken from atomic calculations \cite{haverkort2005spin}. The $F^k$ ($G^k$) integral values representing direct  (exchange) Coulomb interactions are scaled by a factor  $S_1$  ($S_2$) that is usually about 0.8 to account for correlation effects and electronic screening in solids \cite{Groot2005}.
In CMT, the effect of U$_{pd}$ and  
U$_{dd}$ 
are canceled out between ground and excited states since the total number of electrons does not change   
\cite{haverkort2005spin}. 

The XAS spectra can be obtained with the solutions of Eq.~(\ref{eq:ModelHAM}) yielding eigenstates $\ket{\psi_i}$ as a sum of Slater determinants with eigenenergy $E_i$
 \begin{equation}
\hat{H}\ket{\psi_i}=E_i \ket{\psi_i}
\label{eq:Energy}
\end{equation}
Then, the XAS spectrum can be obtained from
 \begin{equation}
I(\omega) =-\frac{1}{Z} \sum_i \Im \left( \braket{\psi_i|\hat{D}^\dagger \frac{1}{\omega -\Hat{H}+E_i + \imath \Gamma/2} \hat{D}|\psi_i} \right)   \exp(-\beta E_i)
\label{eq:spect}
\end{equation}
where $\omega$ represents the excitation energy, and $\hat{D}$ is the dipole operator describing a $p$-electron excitation  into an unoccupied $d$-state. Eq.~(\ref{eq:spect}) results in a Lorentzian broadening of the transition intensities through the imaginary shift $\Gamma$ from the real axis, with $\beta$ being the inverse temperature representing 10 K 
 (if not stated otherwise) 
 and $Z$ is the partition function.   
The broadening of computed transition intensities, frequently performed to facilitate comparison with experimental spectra  \cite{Shariati2013,Bidermane2013a,LucianoTrigueroYiLuo1999,Brena2006,Brena2004a}, employs a Gaussian broadening (GB) for experimental convolution factors additionally to the Lorentzian broadening (LB) for lifetime broadening. GB is characterised by a constant factor, i.e. a full width at half maximum (FWHM). The LB based convolution depends on the core-hole lifetime. Generally, the lifetimes of holes in $p_{3/2}$ and $p_{1/2}$ states differ, leading to differences in  peak widths in the L$_3$ and L$_2$ features, respectively. It is common to employ an energy-dependent stepwise function that uses two broadening factors $\Gamma_1$ for L$_3$ and $\Gamma_2$' for L$_2$ features before and after an energy window defined by $E_1$ and $E_2$, and a linear function within the energy window that starts at $\Gamma_1$ and ends at $\Gamma_2$'  \cite{Bidermane2013a,E.Brumboiu2013}. 
If not stated otherwise, the core hole lifetimes in the tested TM ions were estimated to correspond to values of $\Gamma$ between 0.3 to 0.5 eV with increasing values from Mn to Ni. 
  Gaussian convolution  of 0.1 eV accounted for other effects leading to broadening of the experimental spectra.

 \begin{table*}[!htbp]
\caption{\label{tab:para} Overview of parameters and parameter ranges used to generate the reference dataset, employed in descriptor vector and in output vector of the ANNs, as well as  the sizes of the used datasets. }
\centering
\begin{tabular}{p{4.0cm}p{3.5cm}p{4.5cm}p{4.5cm}}
& A2pX & A2pX-T2 & A2pX-T2EP \\
\hline
\hline
T (K)& 10 & 10 & [1,1000]\\
FWHM (eV) &0.1 &0.1 & [0.05,0.25]\\
\hline
Descriptor vector & [[spectrum]] & [[T,FWHM,spectrum]] & [[T,FWHM,spectrum]]\\
\hline
D$_q$ (eV) & [0.0,0.2]\footnotemark& [0.0,0.2] & [0.0,0.2]\\
D$_s$ (eV) & [-0.2,0.2]& [-0.1,1.4] & [-0.1,1.4]\\
D$_t$ (eV) & [-0.2,0.2]& [-0.6,0.6] & [-0.6,0.6]\\
S$_1$ & [0.60,1.00] & [0.60,1.00] & [0.60,1.00]\\
S$_2$ & [0.60,1.00] & [0.60,1.00] & [0.60,1.00]\\
$\Gamma_1$ (eV)  & element specific\footnotemark & [0.3,0.7] & [0.3,0.7]  \\
$\Gamma_2$=$\Gamma_2$'-$\Gamma_1$ (eV)  & element specific\footnotemark & [0.0,0.2] & [0.0,0.2]  \\
\hline
Output vector &[[D$_q$,D$_s$,D$_t$,S$_1$,S$_2$]] & [[D$_q$,D$_s$,D$_t$,S$_1$,S$_2$,$\Gamma_1$,$\Gamma_2$]]& [[D$_q$,D$_s$,D$_t$,S$_1$,S$_2$,$\Gamma_1$,$\Gamma_2$,\newline J$^2$,J$_z$,S$^2$,L$^2$, \newline
n(a$_{1g}$),n(b$_{1g}$),n(b$_{2g}$),n(e$_{g}$)]] \\
\hline
Reference DSS& 60k& 360k & 360k \\
Tested subset sizes (DSS)  $N$ & [250,2.5k,12.5k,\newline 25k,37.5k] & [50k,100k,200k,300k] & [50k,100k,200k,300k]\\
Training dataset size & 80\% of DSS & 80\% of DSS  & 80\%  of DSS\\
Validation dataset size & 20\% of DSS & 20\% of DSS  & 20\%  of DSS\\
Test dataset size & 2k &10k & 10k \\
\hline
\hline
\end{tabular}
\footnotetext[1]{Except for $C_4$ and O$_h$ symmetry cases in which the D$_q$ ranges were [-0.2,0.2] eV and for the CoPc case with the parameter ranges being increased (see text).}
\footnotetext[2]{Ni:0.48; Fe:0.36; Co:0.43; Mn:0.32 eV}
\footnotetext[3]{Ni:0.52; Fe:0.37; Co:0.43; Mn:0.34 eV}
\end{table*}

\begin{table*}[!htbp]
\caption{\label{tab:anna} Number of neurons in the hidden layers (HL) for  narrow, medium and wide ANN architectures. Layer widths decrease with proximity to the output layer.}
\centering
\begin{tabular}{p{2.0cm}p{3cm}p{4cm}p{5cm}}
& narrow & medium & wide \\
\hline
\hline
 1HL& 16 & 256 & 2048    \\
 3HL& 16;16;8 & 256;256;64 & 2048;2048;1024    \\
 6HL& 16;16;8;8;8;8 & 256;256;64;64;32;32 & 2048;2048;1024;1024;512;512    \\
\hline
\hline
\end{tabular}
\end{table*}

 The solutions of Eq.~(\ref{eq:Energy}) and the consecutive convolution of the transition intensities  
 were obtained with the Quanty code \cite{Haverkort2012} which performs a Lanczos algorithm on a random $d^n$ configuration generated by the tridiagonal Krylov bases to determine the electronic ground state. 

\subsection{Artificial Neural Networks}

In the presented method, an ANN replaces the unknown inverse function, i.e. it can reconstruct  the parameterization of a 2$p$ XAS model Hamiltonian from a spectrum. That is to say; it allows reading the parameterization of Eq. (\ref{eq:ModelHAM})  from a spectrum, including the consecutively applied convolution of transition intensities and external factors.

For accurate results, the  ANN needs a suitable network architecture and effective training, indicated by a positive evaluation of test data, before it can provide reliable results. 
The smallest building block of an ANN is an artificial neuron or node, which is a weighted processing unit with a simple mathematical representation. The activation function $f$ describes how a neuron evaluates given information. Besides widely used sigmoidal functions \cite{Manzhos2015,Majumder2015}, the rectified linear unit (RELU) function is a common choice in ANN supported computational materials science \cite{Goh2017,Sinitskiy2018,Timoshenko2019}. 
Individual neurons are collected in layers. In a fully connected (so-called dense) ANN, each neuron in a layer is connected to all neurons in neighboring layers. This allows passing information in one direction from layer to layer  by so-called forward-propagation. 
In an ANN, a single layer with D$_N$ neurons taking input from $D_i$ neurons in a previous layer is represented as 
\begin{equation} 
y_j= \sum_{p=0}^{D_{N}} w_p f ( \sum_{i=0}^{D_{i}} w_{j_i}x_i +b_p)
\end{equation}
where neurons process the input $x_i$  through the activation function $f$, weights $w_{j,k}$ and bias $b_p$. This process starts at the input layer, i.e. the descriptor vector, and continuous to subsequent hidden layers until it reaches the last layer, the so-called output layer, representing the output vector or feature vector. Then, complex relations can be learned from reference data constructing complex functions through a series of layers and neurons with adaptable weights. 
For multilayer ANNs with $N$ hidden layers, a general form is given by
\begin{equation}
y_j= f_o ( \sum_{k_N=0}^{D_{N}} w_{j,k_N}^{(N)}f_{N-1,k_{N-1}} 
(\sum_{k_N-1=0}^{D_{N-1}} w_{k_N,k_{N-1}}^{(N-1)}f_{N-2,k_{N-2}} 
\times ...( \sum_{k_1=0}^{D_{1}} w_{k_2,k_1}^{(2)}f_{1,k_{1}}
(\sum_{i=0}^{D_i} w_{k_1,i}^{(1)}x_{i}))))
\end{equation}
omitting any bias term. The last activation function $f_o$, which yields the output $y$, employs, in general, a linear function for regression tasks.
  
The ANN architecture describes  the connection patterns between neurons within a network. This affects how neurons pass information through the network. The width is the number of neurons in a layer and the depth refers to the number of layers. Both width and depths are hyperparameters of the ANN in addition to learning rates and, e.g. momentum used in some optimizers.

During training, a randomly selected subset of reference data, i.e. training data, is used to optimize the weights in the ANN through back-propagation employing the gradient of a cost function. The cost function compares and penalizes differences between predicted value ($y_p$), i.e. values computed by the ANN through passing the input vector through the network, and the actual value ($y_a$). Common choices of cost functions for regression tasks are, for instance, the Mean Absolute Error (MAE) or the Root Mean Square Error (RMSE).

Each ANN can be optimized with respect to hyperparameters, which are different from parameters, e.g. the weights, in the ANN.
In contrast to trainable weights of ANNs, hyperparameters must be optimized outside the training of ANNs. A validation dataset containing different data from the training dataset can be used to evaluate the ANN performance concerning hyperparameters besides detecting overfitting.
With adjusted weights and hyperparameters, the ANN is tested on an independent dataset, i.e. test dataset, that was neither used for training nor validation. 
 
\subsection{Artificial Neural Network  Design}

	The input for an ANN is a reference (for training, test and evaluation of the ANN) or an experimental (for application)  spectrum. The output is a set of features that describe the local electronic properties of the central TM ion. Here, three types of ANNs are tested.  The first ANN type, labeled A2pX, can determine d-level positions described by Ballhausen parameters and screening factors  in the model Hamiltonian Eq.~\ref{eq:ModelHAM}, while lifetime broadening factors are taken from literature \cite{retegan_crispy}, the experimental broadening is estimated with an FWHM of 0.1 eV and a temperature of $T=10$ K.  
	
Temperature (T) can influence the appearance of 2$p$ XAS spectra due to the thermal occupation of electronic states at higher energies. The effect is known to lead to noticeable 2$p$ XAS spectral changes in, for instance, Co$^{2+}$ ions, but also Ni$^{2+}$ ions \cite{DeGroot1997,Csiszar2005}. This is expressed in Eq. (\ref{eq:spect}) by the $\beta$ factor leading to a weighted contribution to an XAS spectrum from electronic states different from the ground state.
In the second ANN type (labeled as A2pX-T2), temperature and the experimental convolution by a Gaussian function with a specific FWHM are taken as an external factor and therefore included in the input vector.  In addition,  lifetime broadening factors are part of the output vector, allowing estimations of 2$p$ core hole lifetimes. 

The third ANN type (A2pX-T2EP) can additionally predict expectation values of the 2$p$ XAS MHA, such as the spin state and the occupation of $d$-states in the ground state configuration. This could allow avoiding computing solutions of Eq. (\ref{eq:Energy}).  While the latter is not computationally expensive, including the EP in the ANN has to be explored for  automated analysis and characterization of 2$p$ XAS spectra.
It should be noted, A2pX-T2EP serves two purposes. One is to analyze a spectrum by reconstructing a representative 2$p$ XAS model Hamiltonian. Here, the ANN is said to \textit{analyse} a spectrum in this case. The other purpose is obtaining the same ground-state properties (or expectation values) as a solution to the former Hamiltonian. 
Then, the ANN \textit{predicts} parts of the solutions of the model Hamiltonian, such as the spin state and occupation numbers.

Here, \textit{recomputed} spectra refer to simulated spectra employing the 2$p$ XAS MAH with the parameters determined by an ANN. 
Any difference between recomputed and reference spectrum (and other properties) measures the impact of errors in  parameter estimations on the spectrum. 

For the datasets of A2pX,  reference spectra are computed at 1000 data points in an energy range of 30 eV, including a 5 eV pre-edge window. 
 Datasets containing 60000 spectra were created with randomly generated parameters  for each of the tested TM-ions.  The intervals of the optical parameters were D$_q$=[0.0,0.2], D$_q$=[-0.2,0.0] eV and for O$_h$ and C$_4$ CFs, respectively, and D$_q$=[0.0,0.2], D$_t$=[-0.2,0.2], and D$_s$=[-0.2,0.2] eV for D$_{4h}$ CFs. The screening parameters were between 0.6 and 1.0. 
The features of the descriptor vector are reference spectra. The optical and screening parameters are the features of the target vector to be determined by the ANNs.
 
  For A2pX-T2(EP), separate databases are required that contain additionally  information needed for the input and output vectors.
The  extended databases include T from 1 to 1000 K, the FWHM of the Gaussian broadening in a range from 0.05 to 0.25 eV. The lifetime broadening factors intervals are [0.3,0.7] eV for $\Gamma_1$ and  [0.0, 0.2] eV with $\Gamma_2$ with $\Gamma_2$' = $\Gamma_1$ + $\Gamma_2$. 
The input vector size increases by two features, which are T and FWHM, and the output vector size increases from 5 to 15 features for A2pX-T2EP. The output vector includes lifetime broadening factors for L$_2$ and L$_3$, expectation values for $J^2$, $J_z$, $L^2$ and $S^2$, the occupations of a$_{1g}$, b$_{1g}$, b$_{2g}$ and e$_g$ states, besides the already introduced parameters used in the simpler ANN (A2pX) that are the screening factors S$_1$ and S$_2$, and the description of the $d$-level splitting through D$_q$, D$_s$ and D$_t$.  The increasing number of features in the output vector  can decrease the accuracy of the ANN due to the increased complexity of the learning problem \cite{Huo2017}. 
Hence, the number of reference spectra was also increased.  Databases containing 360000 spectra were created for each of the tested TM ions, from which 10000 spectra were randomly selected for testing.  Details are given in Tab.~\ref{tab:para}.

\subsection{Dataset split and training details}

Before ANN training, a dataset of a TM in a CF symmetry was split into three different subsets for training, validation and testing.  Besides, all features of the input and output vectors, i.e. parameters and spectra, were normalized in [0,1] intervals (from here on referred to as box normalization). 
For A2pX, for instance, first 2000 spectra were randomly selected as test datasets. 
Then, the remaining data (not including the test dataset) was used to construct several subsets with sizes of $N$ between 250 and 35000 spectra. This allows probing the required dataset size (DSS) for ANN training. 
 The latter subset were randomly split into training and validation datasets with a ratio of 80\% to 20\%, respectively.
For A2pX-T2(EP), $N$ ($N<=350000$) of the remaining spectra were randomly selected and also divided into training and validation datasets with a ratio of 80:20. This follows the same approach as for the A2pX dataset splitting. 
 An overview of the computational setup and used parameters for these cases is given in the first data column of Tab.~\ref{tab:para}.

The validation data were not used to train the model, but to monitor the training progress, including the detection of over- and underfitting by comparing the evolution of the cost function values of training and validation datasets.  Furthermore, the validation datasets were employed to optimize and validate hyperparameters. The tested hyperparameters include the initial learning rates, choice of the neuron activation functions  in each layer, as well as layer width and ANN depth. A grid search was undertaken to find optimal learning rates and a Monte Carlo search for different combinations of activation functions in the HLs that included sigmoidal functions, the exponential linear unit and the RELU. 
As a result of the grid search, the RELU activation function was used in the input and each hidden layer because no other combination of activation functions yielded equally accurate or  significantly improved results.
The tested ANNs consisted of  fully connected  layers of neurons.  A manual search was performed with nine selected cases with different depths and widths and ANN architectures included one, three and six hidden layers. The number of neurons in the hidden layers is given in Tab.~\ref{tab:anna} for narrow, medium and wide ANNs. Details on input and output layer sizes and features are given in Tab.~\ref{tab:para}.  
All ANNs were implemented with the high-level neural network API Keras \cite{Chollet2015}.  
A stochastic gradient descent method \cite{Zeiler2012} iteratively optimized the weights of the ANNs in 250 epochs in the training phase.
The learning rates were reduced by a factor of 0.2  to fine-tune the optimization when ten consecutive iterations (so-called epochs) did not result in improvement of the cost function value taken as MAE between y$_a$ and y$_p$.
In addition, the mean squared error (MSE) and the RMSE are reported.
  
\section{Results and discussion}

In the following, the capabilities of ANNs applied to spectra of Co$^{2+}$, Ni$^{2+}$, Fe$^{2+}$ and Mn$^{2+}$ central ions are discussed. Details are presented for Co$^{2+}$ (if not stated otherwise), while further information about  other TM ions is in the Supporting Information (SI).

\subsection{Training, Validation and Testing}
\begin{figure}[!htbp]
\centering
\includegraphics[width=1.0\textwidth]{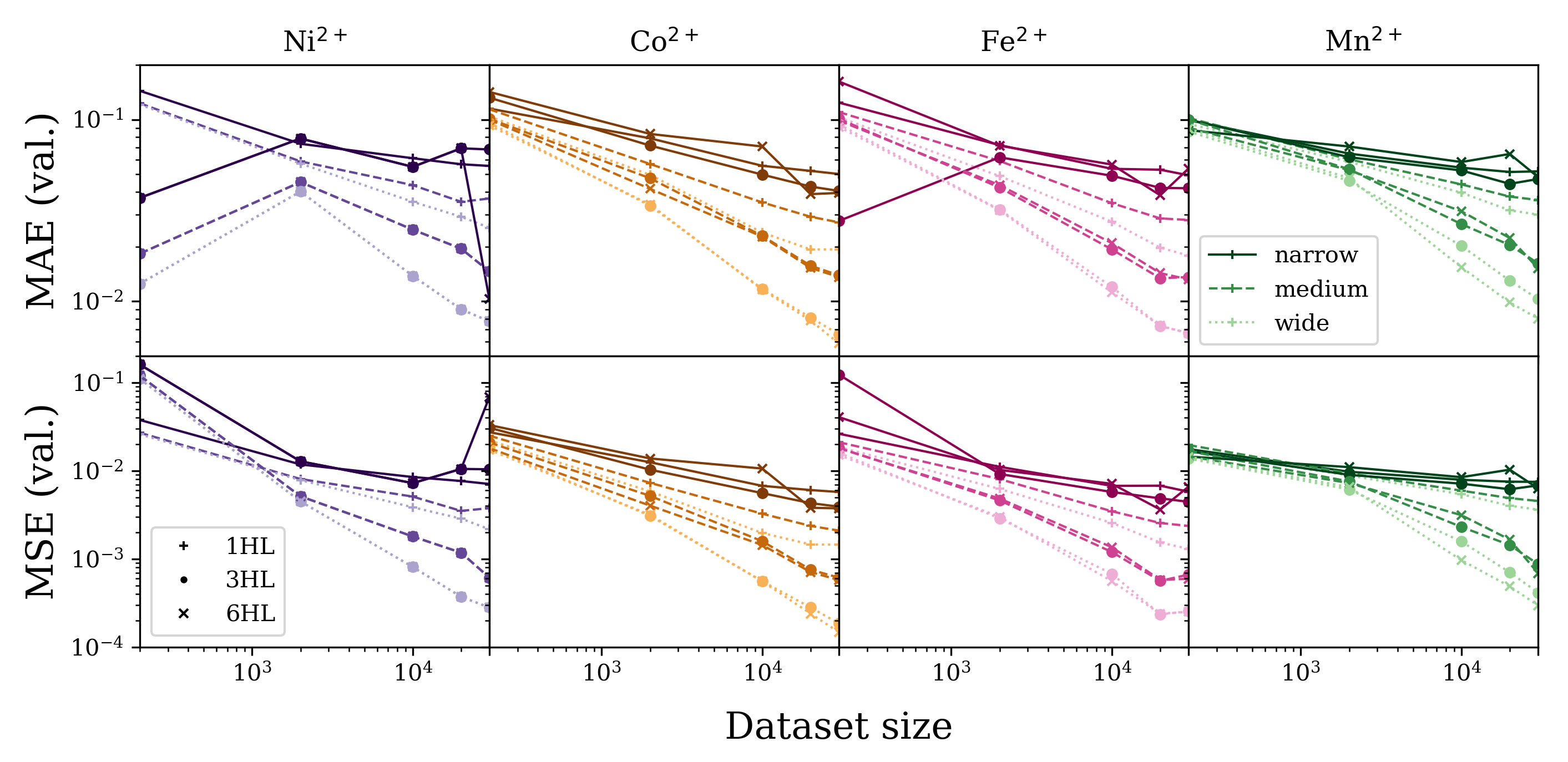}
\caption{(Color online)  
The MSE and MAE of validation data on trained ANNs with different architectures with one, three and six HLs for (from left to right) Ni$^{2+}$, Co$^{2+}$, Fe$^{2+}$ and Mn$^{2+}$ in D$_{4h}$ CF as a function of dataset size.
}
\label{fig:tre}
\end{figure}

Figure~\ref{fig:tre} shows the MAE and the MSE for the trained ANNs with one, three and six hidden layers for different training dataset sizes. The figure shows the results for Ni$^{2+}$, Co$^{2+}$, Fe$^{2+}$ and Mn$^{2+}$ ions in a CF with D$_{4h}$ symmetry. 
 MAE and  MSE decrease with increasing dataset size. There is little to no improvement for ANNs with only a few neurons or for narrow ANNs. Wide ANN architectures with one hidden layer can already achieve some agreement between predicted and actual data. Comparatively, the MSE drastically increased for Ni$^{2+}$ with only one HL at large dataset sizes. For most cases of ANNs consisting of several hidden layers, training on datasets larger than 20000 spectra yielded  very accurate predictions of  optical and screening parameters.
MSEs of less than 10$^{-3}$ (MAE of less than 10$^{-2}$) were achieved for the medium and wide ANNs. 
The analysis of the trained ANNs for  O$_h$ and C$_4$ CFs 
(shown in Fig.~\ref{fig:tre_oh})  results in similar trends that are: i) often accuracy increases  with increasing training dataset size and ii) deeper ANNs with more neurons (i.e. wider) obtain smaller MSE/MAE.
For O$_h$ symmetry, the reduced parameter space causes MSE and MAE  to be be approximately  one order of magnitude smaller at large dataset sizes.  In general, the results suggest to use wide ANN with three to six HLs - the latter will be the basis for the remaining discussion.

\begin{figure}[!htbp]
\centering
\includegraphics[width=1.0\textwidth]{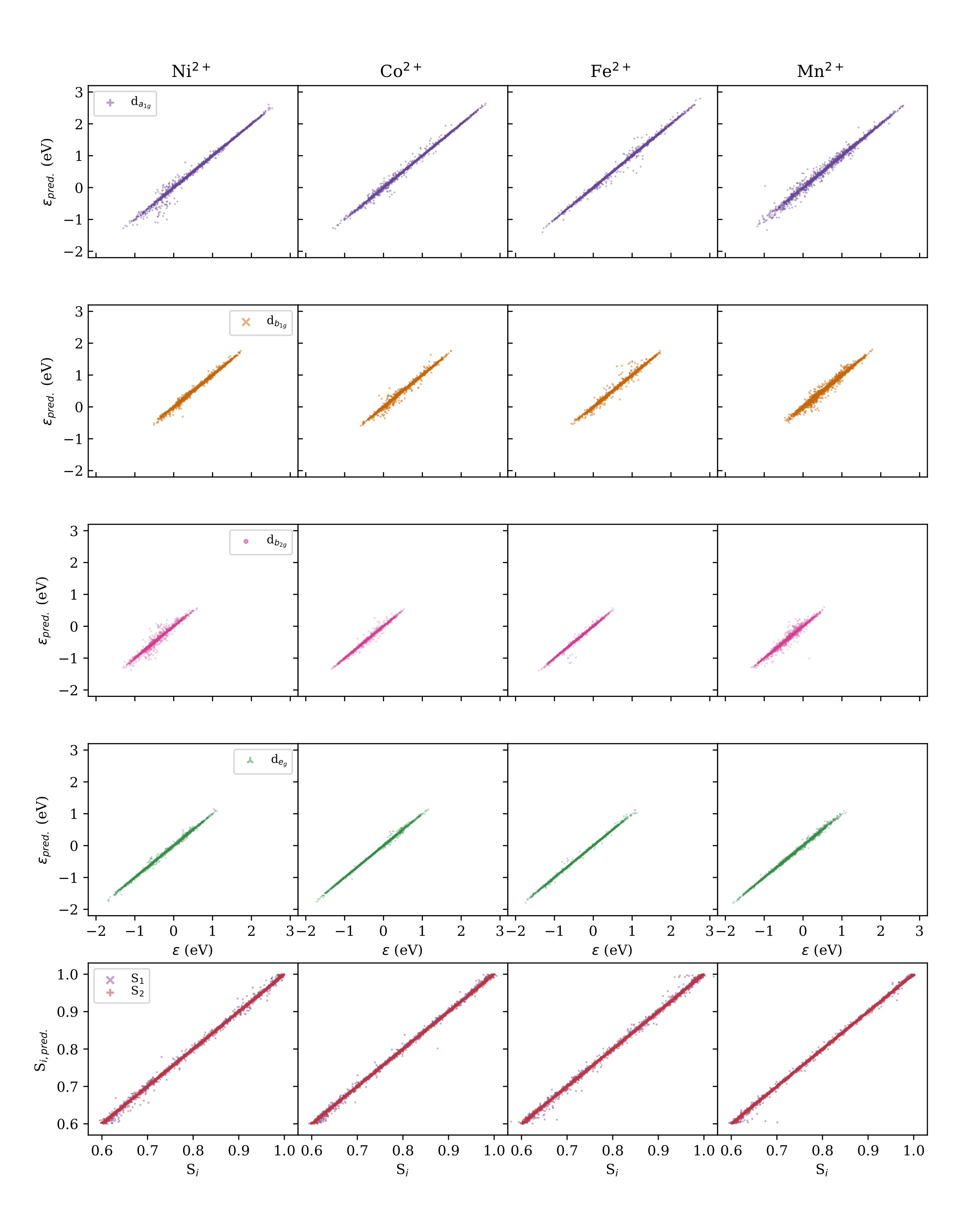}
\caption{(Color online)  
From top to bottom: Actual vs predicted (pred.) relative positions of $d$-orbital energies $\epsilon$ for d$_{a1g}$, d$_{b1g}$, d$_{b2g}$, d$_{eg}$  and screening parameters (S$_1$ and S$_2$), and from left to right: Ni$^{2+}$, Co$^{2+}$, Fe$^{2+}$ and Mn$^{2+}$ in wide ANNs with six HLs. Data are computed with the validation dataset on the trained ANNs (A2pX). 
}
\label{fig:dstates}
\end{figure}

	Figure \ref{fig:dstates} shows a comparison of the actual and the predicted orbital energies (evaluated with the predicted Ballhausen parameters) as well as the screening parameters of the Slater-Condon integrals. The results are given  for Ni$^{2+}$, Co$^{2+}$, Fe$^{2+}$ and Mn$^{2+}$ in a D$_{4h}$ CF. The orbital energies have a range from -1.8 eV to 2.8 eV. 
For all tested elements,   the ANNs accurately detect  the screening parameters.  
However, there are noticeable differences in  accuracy of  orbital energies. Energy predictions of d$_{b_{1g}}$ and d$_{b_{2g}}$ are less accurate than for d$_{a_{1g}}$ and d$_{e_{g}}$ states. This is particularly obvious for d$_{b_{2g}}$ in Ni$^{2+}$ and for d$_{b_{1g}}$ in Mn$^{2+}$. The effect is strongest for orbital energies close to 0 eV, suggesting that those computed spectra are similar to each other and that the contributions from d$_{b_{1g}}$ and d$_{b_{2g}}$ states to these XAS spectra are not significantly different from  being detected with high accuracy. On the other hand, d$_{a_{1g}}$ and d$_{e_{g}}$ are more accurately determined for all tested elements. 

Table~\ref{tab:res01} provides  MAEs and RMSEs obtained with A2pX (wide and six HL) on the test data for the  TM ions in  D$_{4h}$ symmetric CFs. The MAE (less than 2\%) and RMSE (less than 5\%) are of similar magnitude but slightly larger than those given in Fig.~\ref{fig:tre} for the validation data demonstrating effective training and potentially useful ANNs. Large errors are in D$_q$ and D$_s$ (RMSEs of 1.0 to 4.5\%), while the screening parameters and D$_t$ results are more accurate (0.8 to 2.4\%). A corresponding table (Tab. \ref{tab:res_A2pX}) for O$_h$ symmetry  is in SI. The errors for this case are smaller  (less than 0.5 \%), reflecting a decrease in complexity by a reduced parameter space, which also benefitted the training results.
 
  \begin{table*}[!htbp]
\caption{\label{tab:res01}  RMSE (MAE) of normalized D$_q$, D$_s$, D$_t$, S$_1$ and S$_2$ determined on the normalized test datasets (2000 spectra) for Mn$^{2+}$, Fe$^{2+}$, Co$^{2+}$ and Ni$^{2+}$ with A2pX (wide six HL). Values given in percent.}
\centering
\begin{tabular}{p{2.0cm}p{3cm}p{3cm}p{3cm}p{3cm}}
\hline
\hline
& Mn$^{2+}$  & Fe$^{2+}$ & Co$^{2+}$& Ni$^{2+}$ \\
\hline
D$_q $ & 4.28 (1.96) & 2.52 (1.25)  & 3.77 (1.86) & 2.54 (1.29) \\
D$_s$ & 4.48 (1.72)  & 3.39 (1.44) & 0.95 (0.57) &  3.68 (1.53)\\
D$_t$ & 2.44 (1.20)  & 1.77 (0.89) & 1.19 (0.75) & 1.80 (0.94)\\
S$_1$ & 1.30 (0.76)  & 1.19 (0.75) &  2.43 (1.26) & 1.18 (0.77)\\
S$_2$ & 0.84 (0.61)  & 0.86 (0.59) & 1.35 (0.87) & 0.68 (0.48)\\
\hline
\hline
\end{tabular}
\end{table*}

  \begin{table*}[!htbp]
\caption{\label{tab:res02}  RMSE (MAE) of normalized D$_q$, D$_s$, D$_t$, S$_1$, S$_2$, $\Gamma_1$ and $\Gamma_2$ determined on the normalized test datasets (10000 spectra) for Mn$^{2+}$, Fe$^{2+}$, Co$^{2+}$ and Ni$^{2+}$ with A2pX-T2.  Values given in percent.}
\centering
\begin{tabular}{p{2.0cm}p{3cm}p{3cm}p{3cm}p{3cm}}
\hline
\hline
& Mn$^{2+}$  & Fe$^{2+}$ & Co$^{2+}$& Ni$^{2+}$ \\
\hline

D$_q$ & 2.86 (1.71) & 2.76 (1.53) & 1.93 (1.14)& 2.69 (1.37)\\
D$_s$   & 1.11 (0.73) & 1.37 (0.84) & 1.59 (1.01) & 3.01 (1.55)\\
D$_t$ &  0.97 (0.63) & 1.06 (0.69) & 1.35 (0.83) & 2.37 (1.36) \\
S$_1$ & 1.83 (1.20)  & 2.00 (1.33)  & 2.43 (1.60) & 2.21 (1.44)\\
S$_2$  & 1.46 (1.00) & 1.48 (1.02) & 1.48 (1.01) & 1.19 (0.83)\\
$\Gamma_{1}$& 2.83 (1.94)  & 2.84 (1.91) & 1.95  (1.39) & 1.39 (0.98)\\
$\Gamma_{2}$  & 8.63 (5.64) & 7.96 (5.13) &  6.88 (4.18) & 5.77 (3.53) \\
\hline
\hline
\end{tabular}
\end{table*}
 
For A2pX-T2 and A2pX-T2EP,  the same generic hidden layer architecture, as used in A2pX containing up to six dense HL, can obtain comparable accuracy. Initial tests indicated that modifications in training, architecture design and optimization approaches lead to similar outcomes as those seen by A2pX.  
  The common trends are: i) wide and deep (six HL) ANNs perform better than narrow and flat (one HL and three HL) architectures, ii) RELU as the activation function for all layers provides similar or better accuracy than different combinations of other activation functions of the HL and iii) a small number of training data  can limit the accuracy. 
 However, an increased number of training data is required to achieve similar accuracy between A2pX and A2pX-T2(EP). The number of training data increased from ca. 5000 spectra per feature in the output vector in A2pX to more than 15000 spectra per feature.  This reflects  the increased complexity the method faces when T and the FWHM  are part of the descriptor vector.
   
\begin{figure}[!htbp]
\centering
\includegraphics[width=0.95\textwidth]{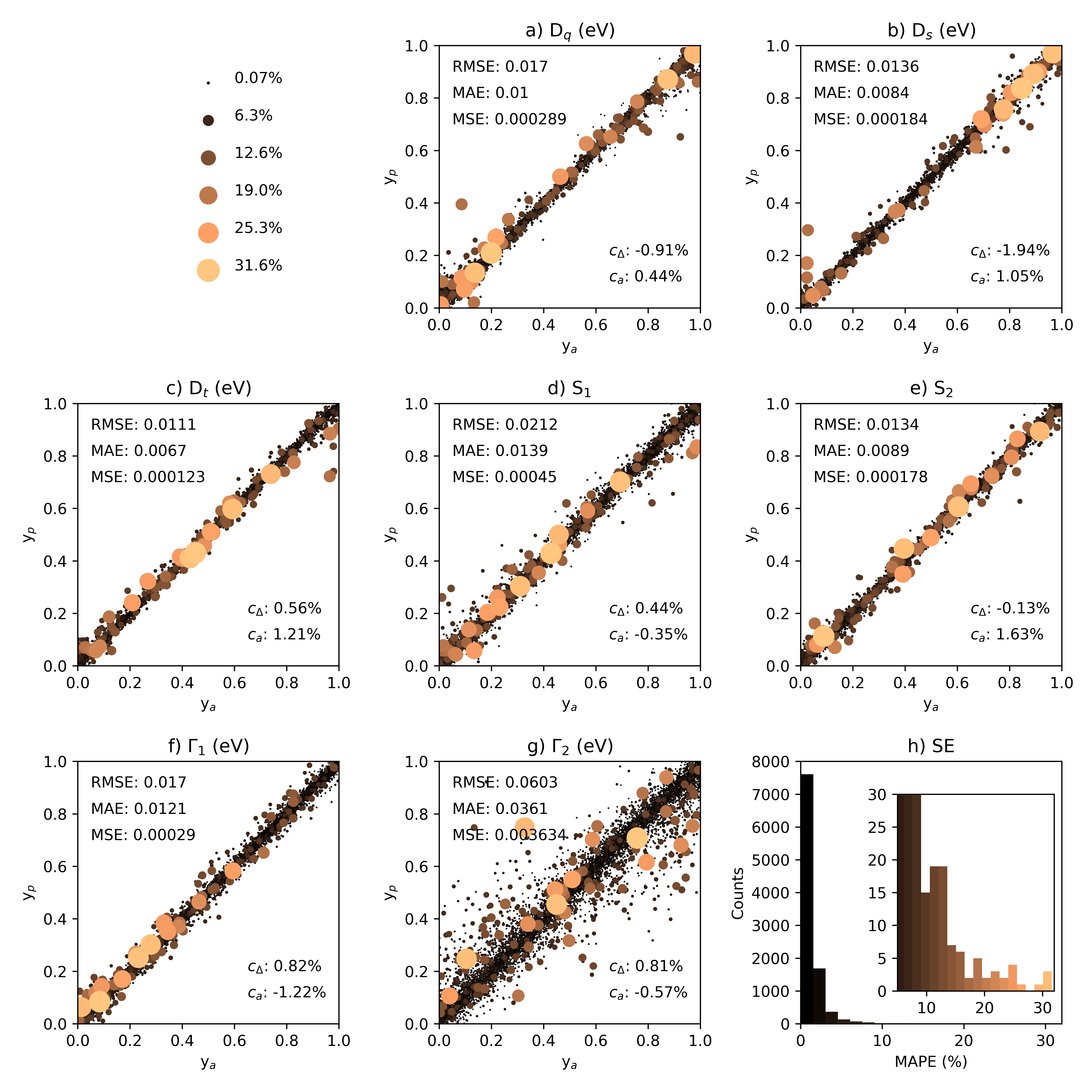}
\caption{(Color online) From a) to g): Comparison of normalized $y_p$ and $y_a$ in D$_q$, D$_s$, D$_t$, S$_1$, S$_2$ $\Gamma_1$ and $\Gamma_2$ for  Co$^{2+}$ with variable T and FWHM in the description vector evaluated on the test dataset (10000 spectra). Data point sizes and shades represent the SE between recomputed and reference spectra. Figure h) shows the distribution of SE in a histogram in which the color shades are defined for a) to g). The insert in h) shows a part of the histogram with larger resolution in the counts axis. }
\label{fig:copc_tier2}
\end{figure}

 Table~\ref{tab:res02} gives a summary of  RMSEs and MAEs  for A2pX-T2. Performance details  of A2pX-T2EP are in Tab.~\ref{tab:res03}, for which Fig.~\ref{fig:copc_tier2} presents corresponding details  for Co$^{2+}$ and Figures~\ref{fig:ni_ael}, \ref{fig:fe_ael} and \ref{fig:mn_ael} in the SI show the data for Ni, Fe and Mn for D$_{4h}$ cases, respectively. 
For the Co$^{2+}$ ion, the RMSE values are less than 5\%  for all parameters but for $\Gamma_2$ (6.0\%). The average RMSEs for parameter determination (2-3\%) are comparable for A2pXT2 and A2pX-T2EP.  The ANN determines D$_t$ and D$_s$ more accurately than D$_q$. Besides  $\Gamma_2$, S$_1$ also has a relatively large error with ca. 2\%. Regarding EP, the ANN yields RMSE of less than 2\% for all expectation values but $<J_z>$ (2.4\%).
  For the other ions, the trends are similar to those seen for Co$^{2+}$.  
Concerning the determination of the 2$p$ XAS parameters, the RMSEs and MAEs among the tested elements are comparable but, in general, larger for Ni and Fe than for Co. In systems with O$_h$ symmetry, this trend is similar, but  errors are generally smaller, due to reduced complexity.  Details  are given  in Tab.~\ref{tab:res02} and \ref{tab:res03} for D$_{4h}$ CF and in Tab.~\ref{tab:res_oh_A2pXt2} and \ref{tab:res_oh_A2pXt2ep} for  O$_h$/C$_v$ CF for A2pX-T2 and Ap2X-T2EP, respectively. However, individual output features can show significantly larger deviations.     
Most significant RMSEs are seen in $\Gamma_2$, that is 5 to 9\% in A2pX-T2(EP)  followed by either $\Gamma_1$ (1 to 3\%) or S$_1$ (1.7 to 2.4\%) and depend on element and CF symmetry. 
Among  Ballhausen parameters in D$_{4h}$ systems, the largest errors are in D$_q$. For A2pX, the RMSEs are between 3 and 5\%, while they are 2 to 3\% for A2pX-T2(EP).
The RMSEs of other parameters are smaller, which confirms a relative insensitivity of computed spectra regarding small changes of D$_q$ as pointed out by previous studies \cite{DeGroot1997}.
 
For a quantitative analysis of ANN performance with more features in the output  vector, i.e. A2pX-T2 and A2pX-T2EP, an additional measure for quantifying errors in the output vector  associated with spectral features is introduced. Here, the mean absolute percentage error (MAPE) of recomputed spectra is taken with respect to the normalized superposition of the recomputed and reference spectrum. The MAPE, here called the spectrum error (SE), is employed to relate the influence of feature errors  in normalized output vector ($y_p -y_a$), labeled as $\Delta$, in the corresponding spectra.  SEs are evaluated for all spectra in  test datasets and their corresponding recomputed spectra. Examples comparing recomputed and test spectra for A2pX-T2EP covering the whole range from smallest to largest SE, are given in Fig.~\ref{fig:ni_comp}, \ref{fig:fe_comp}, \ref{fig:co_comp} and \ref{fig:mn_comp}  for  Ni$^{2+}$, Fe$^{2+}$, Co$^{2+}$ and Mn$^{2+}$ (D$_{4h}$), respectively. The figures compare test spectra (black curve) with recomputed spectra  (dashed red) and are ordered according to their SE.
 
By visual inspection, a SE of less than 3\% gives excellent agreement, while SE of 4 to 7\% seems still acceptable to some degree. At larger values, differences between recomputed and reference spectra can become significant. This can be observed in, e.g., additional/missing features that appear more frequently. At SE of 10\% and larger values, mismatches in the spectra and overall different spectral shapes are common. In some cases, a small shift, for instance, can cause SEs between 10 and 16\%. 
For all tested ions, ca. 90\% of the recomputed test spectra have a SE of less than 3\%. The 7\% threshold is reached for 97.6, 98.4, 98.4 and 97.6\% of all test spectra for Ni$^{2+}$, Mn$^{2+}$, Fe$^{2+}$ and Co$^{2+}$, respectively.
The largest SE is  noticeable larger for Ni$^{2+}$ (34\%) than for other elements (ca. 30\%), while Mn$^{2+}$ has a max. SE of only 24\%. This demonstrates that for simulated 2$p$ XAS signals, the ANNs can  reliably  reproduce parameterizations.

The main purpose of extending A2pX-T2 to A2pX-T2EP is to estimate expectation values of electronic states such as momentum and orbital occupations. As seen above, A2pX-T2EP and A2pX-T2 obtain similarly accurate outcomes for the test spectra. Hence, both are equally capable of determining the parameterization of the 2$p$ XAS model Hamiltonian. Hence, the following discussion will focus on A2pX-T2EP.

 For A2pX-T2EP, Figure~\ref{fig:copc_tier2} shows the predicted ($y_p$) and actual values ($y_a$) of  Ballhausen parameters, as well as screening and broadening factors at variable temperature and FWHM evaluated for the test dataset in normalized form for Co$^{2+}$.     The distribution of SEs is given in histograms in Fig.~\ref{fig:copc_tier2} for Co$^{2+}$ and in Fig. \ref{fig:ni_ael}, \ref{fig:fe_ael} and \ref{fig:mn_ael} for Ni$^{2+}$, Fe$^{2+}$ and Mn$^{2+}$, respectively. The latter three figures contain also the distribution of $y_p$ and $y_a$. The 21 bins in each histogram are the basis for sizes and shading of  data points in the other subplots. In addition, the Pearson correlation between SE and the residuals ($\Delta$), given as $c_{\Delta}$, or the relative values of $y_a$, given as $c_a$, are indicated for each parameter.  The correlation factors are small and reach only in a few cases, values being marginally larger than 2\%: that is S$_1$ for Fe and D$_t$ for Mn ions.  Overall, the results show acceptable accuracy for the majority of test cases. Comparing y$_p$ and y$_a$, only a few SE outliers  (i.e. large SE values) are present and most of them are very close to the diagonal. The largest SE in the Co$^{2+}$ test dataset is partly due to a miscalculation of $\Gamma_2$.  Given also these cases, it supports that an output vector with several small feature errors can cause large SEs.

Among the tested ions, the D$_q$ distributions of actual vs. predicted data points   have a slightly larger number of significant outliers among  Ballhausen parameters (also given by larger RMSE / MAE values). In contrast, there is a somewhat wider spread of data points near the diagonal for D$_s$. $\Gamma_2$  is difficult to determine for all ANNs and shows therefore the widest distributions.
The plots  reveal that, in some cases, data points with large SEs are forming clusters in certain $y_a$ ranges. Clustering in positive/negative regions of y$_a$ correlates with negative/positive values of c$_a$.  At small values of D$_q$, S$_1$ and $\Gamma_1$ or large values of D$_s$,  differences between recomputed and reference spectra increase for Co$^{2+}$.  For the Mn$^{2+}$ and Fe$^{2+}$ ions, there are clusters of large SE in D$_t$. Other parameters and the results for Ni, in general, show weaker indications of SE clusters. 

Note that in some cases, atomic parameters can have large $\Delta$s. Despite this disagreement, those data points  show a reasonable resemblance between the recomputed and reference spectrum (SE of less than 7\%) in many instances. This is seen, for instance, in D$_q$ and D$_t$ for Ni$^{2+}$ in Fig.~\ref{fig:ni_ael}.  Hence, not every single-parameter estimation error results in  large  SEs.   In contrast, spectra with large  SEs can be very close to the diagonal. This implies that  reasonably accurate estimates of most parameters still can result in significant SEs for a few instances.  This intrinsically limits the accuracy of the ANN-based method.  
In contrast, large mismatches in single or few features do not necessarily yield in large SEs. 
Consequently,  small  changes in spectra could result in noticeable $\Delta$s for single or a few parameters in a limited number of cases. 

 \begin{figure}[!htbp]
\centering
\includegraphics[width=0.95\textwidth]{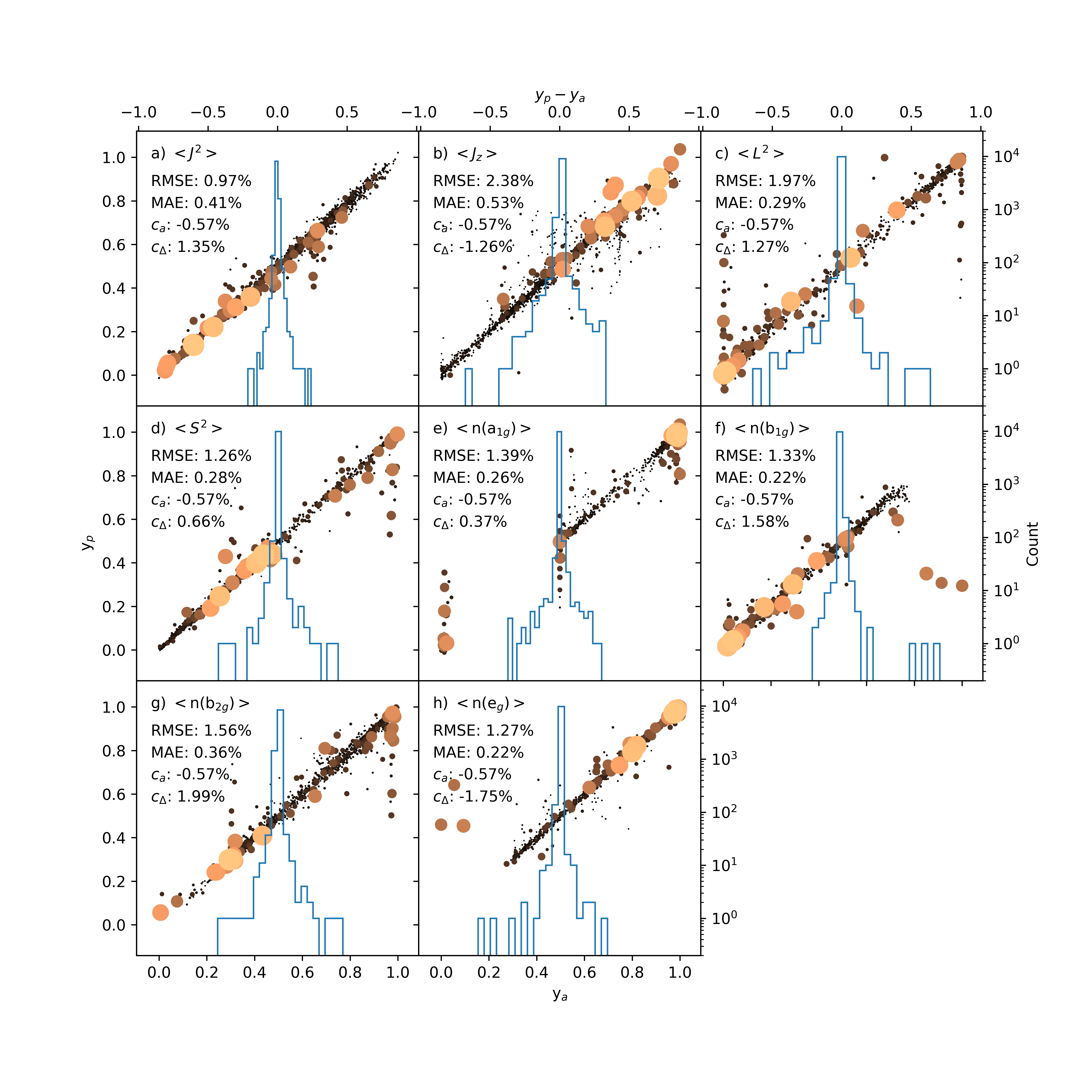}
\caption{
(Color online) Comparison of normalized $y_p$ vs $y_a$ for predicted properties by A2pX-T2EP for Co$^{2+}$ determined on test dataset (10000 spectra). The overlayed histograms show the distribution of  difference between $y_p$ vs $y_a$ on a logarithmic count scale.  Color code and data point sizes are as in Fig.~\ref{fig:copc_tier2}. The scales of the histograms are on the top right, the scales of predicted (y$_p$) vs actual data (y$_a$) are given on the bottom  left. Pearson correlation to SE for $\Delta_y =y_p - y_a$ (c$_{\Delta}$) and y$_a$ (c$_a$) are given. }
\label{fig:copc_tier2_el}
\end{figure}

Figure~\ref{fig:copc_tier2_el} compares $y_p$ and y$_a$ of selected EP, including $J^2$, $J_z$, L$^2$, S$^2$, and d-state occupations in a Co$^{2+}$ ion in  D$_{4h}$ CFs. The same datapoint scale as in Fig.~\ref{fig:copc_tier2} applies. In addition, histograms of the  residuals on a logarithmic count scale are overlayed.  Figures~\ref{fig:ni_pel}, \ref{fig:fe_pel} and \ref{fig:mn_pel} in SI show the results for Ni$^{2+}$, Fe$^{2+}$ and Mn$^{2+}$, respectively. The corresponding RMSEs and MAEs are summarised in Tab.~\ref{tab:res03} for D$_{4h}$ and in Tab.~\ref{tab:res_oh_A2pXt2ep} for O$_{h}$/C$_{4v}$ systems.  
  Overall, the errors are comparable but slightly larger (with the exception of the broadening factors) for EP than for analyzed parameters.  This is partly caused by a few data points with very large $\Delta$s, which can be 50 to 100\%, for a few EP features.  RMSEs of less than 2\% were yielded for $<J^2>$, $<L^2>$ and $<S^2>$. Prediction of J$_z$ expectation values shows the largest uncertainties in all cases, this includes elements as well as ligand-field symmetry.   For this parameter, the RMSEs are between 2 to 5\% for D$_{4h}$ and 2 to 11\%  for O$_h$ ligand-fields. 
Values of orbital occupations, in general, show smaller RMSEs, whereas the occupation of  t$_{2g}$ and e$_g$ states in Ni$^{2+}$ (O$_h$) reaches RMSEs of almost 10\%. 
 This error is less for other elements, e.g. ca. 4\% in Co$^{2+}$  O$_h$, and less than 3 \% in D$_{4h}$ systems. 
  Clustering of data points with large SE is also pronounced for EP. In Co$^{2+}$ D$_{4h}$, for instance, $<J^2>$, $<S^2>$ and n(b$_{1g}$) contain clusters  for small values while $<J_z>$ and n(e$_g$)  have them for large values. 
Large SE values have a wider spread in the expectation value prediction than for parameterization.
 Still, distributions in histograms show that  expectation value estimation is accurate in most cases, i.e., the majority of $\Delta$s is in an acceptable range.  However, the observed  large uncertainties (close to 100\%) in a few instances might limit  applications for A2pX-T2EP, because  there is no absolute certainty that the presented approach will yield correct electronic configurations. 
Interestingly,  prediction of expectation values benefits from passing the spectra in the descriptor vector. In contrast, another ANN architecture (dense  six HL in the wide setup) did not obtain reliable results (i.e. large RMSE and MAE) when the descriptor vector contained only the parameters from the model Hamiltonian. This included two cases; one for which the parameters were determined by A2pX-T2(EP) and the other one for which the parameters were taken directly from the test dataset. 

\begin{table*}[!htbp]
\caption{\label{tab:res03} RMSE (MAE) of normalized D$_q$, D$_s$, D$_t$, S$_1$, S$_2$, $\Gamma_1$ and $\Gamma_2$, and predicted the electronic configuration determined on the test datasets for Mn$^{2+}$, Fe$^{2+}$, Co$^{2+}$ and Ni$^{2+}$ with A2pX-T2EP.  Values given in percent.}
\centering
\begin{tabular}{p{2.0cm}p{3cm}p{3cm}p{3cm}p{3cm}}
& Mn$^{2+}$ & Fe$^{2+}$  & Co$^{2+}$& Ni$^{2+}$ \\
\hline
\hline
D$_q$  	& 2.95 (1.71) 		& 2.10 (1.27) 	& 1.70 (1.00) 		& 2.75 (1.29)\\
D$_s$  	& 0.99 (0.67) 		& 1.08 (0.70) 	& 1.36 (0.84) 		& 3.07 (1.52)\\
D$_t$  	& 0.94 (0.60) 		& 1.00 (0.59) 	& 1.11 (0.67) 		& 2.46 (1.34)\\
S$_1$ &    1.73 (1.16)			& 1.80 (1.19) 	& 2.12 (1.39)		& 2.27 (1.40)\\
S$_2$ & 	1.40 (0.93) 			& 1.40 (0.93)  & 1.33 (0.89) 		& 1.21 (0.80)\\
$\Gamma_{1}$   	& 2.60 (1.76) &  2.49 (1.65)	&  1.70 (1.21) 	& 1.33 (0.92)\\
$\Gamma_{2}$	& 7.94 (5.09) &  7.14 (4.48) 		& 6.03 (3.61)			& 5.34 (3.19)\\

\hline
$<$J$^2$$>$ & 1.49 (0.35) 	&1.46 (0.39) & 0.93 (0.39) & 1.35 (0.34)\\
$<$J$_z$$>$ & 4.93 (1.11) 	& 3.14 (0.44) & 2.35 (0.52) & 3.05 (0.31)\\
$<$S$^2$$>$ & 1.49 (0.23) 	& 1.22 (0.25) & 1.97 (0.29) & 1.65 (0.21)\\
$<$L$^2$$>$ & 1.42 (0.23) 	& 1.17 (0.24) & 1.26 (0.28) & 1.57 (0.27)\\
n(a$_{1g}$) & 1.36 (0.23)  & 1.74 (0.33) & 1.39 (0.26) & 1.68 (0.19)\\
n(b$_{1g}$) & 1.83 (0.38) & 1.18 (0.24) & 1.30 (0.21) & 1.04 (0.28)\\
n(b$_{2g}$) & 1.25 (0.26) & 1.69 (0.29) & 1.51 (0.35) & 2.11 (0.44)\\
n(e$_{g}$) & 0.80 (0.16) 	& 0.96 (0.18) &  1.25 (0.22) & 1.30 (0.25)\\
\hline
\hline
\end{tabular}
\end{table*}
 
In the end, the ANN-based analysis of 2$p$ XAS spectra and a simultaneous prediction of electronic configurations are obtainable to high accuracy for most spectra, while errors in electronic property prediction can be substantial in a limited number of cases.  The most pronounced increase in RMSE/MAE among the different types of ANN (A2pX, A2pX-T2 and A2pX-T2EP) was observed in the screening factors and $\Gamma_2$. Besides, shifts in spectra can also cause large SEs. In addition, ANNs can include effects of temperature and experimental broadening as an additional input feature to accurately estimate the parameters of the 2$p$ XAS model Hamiltonian.

\subsection {Impact of Noise and Background Signal}

 \begin{figure}[!htbp]
\centering
\includegraphics[width=0.95\textwidth]{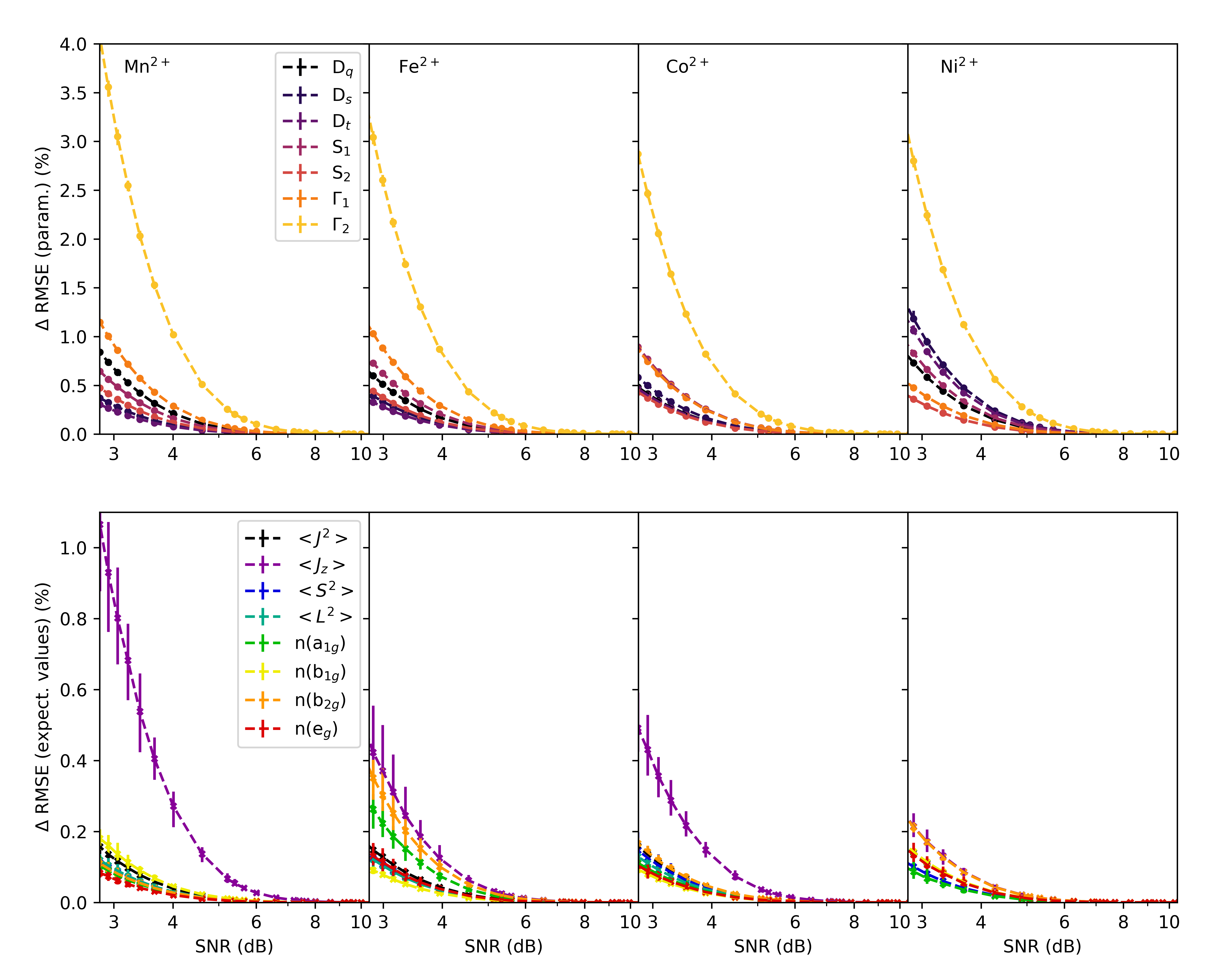}
\caption{(Color online) Influence of noise in test spectra on the ANN's analytic (top: parameter estimation) and prediction (bottom: expectation value estimation) performance  (A2pX-T2EP) evaluated as RMSE between $y_p^{noise}$ and $y_a$.  }
\label{fig:rmse_noise}
\end{figure}

  \begin{figure}[!htbp]
\centering
\includegraphics[width=0.95\textwidth]{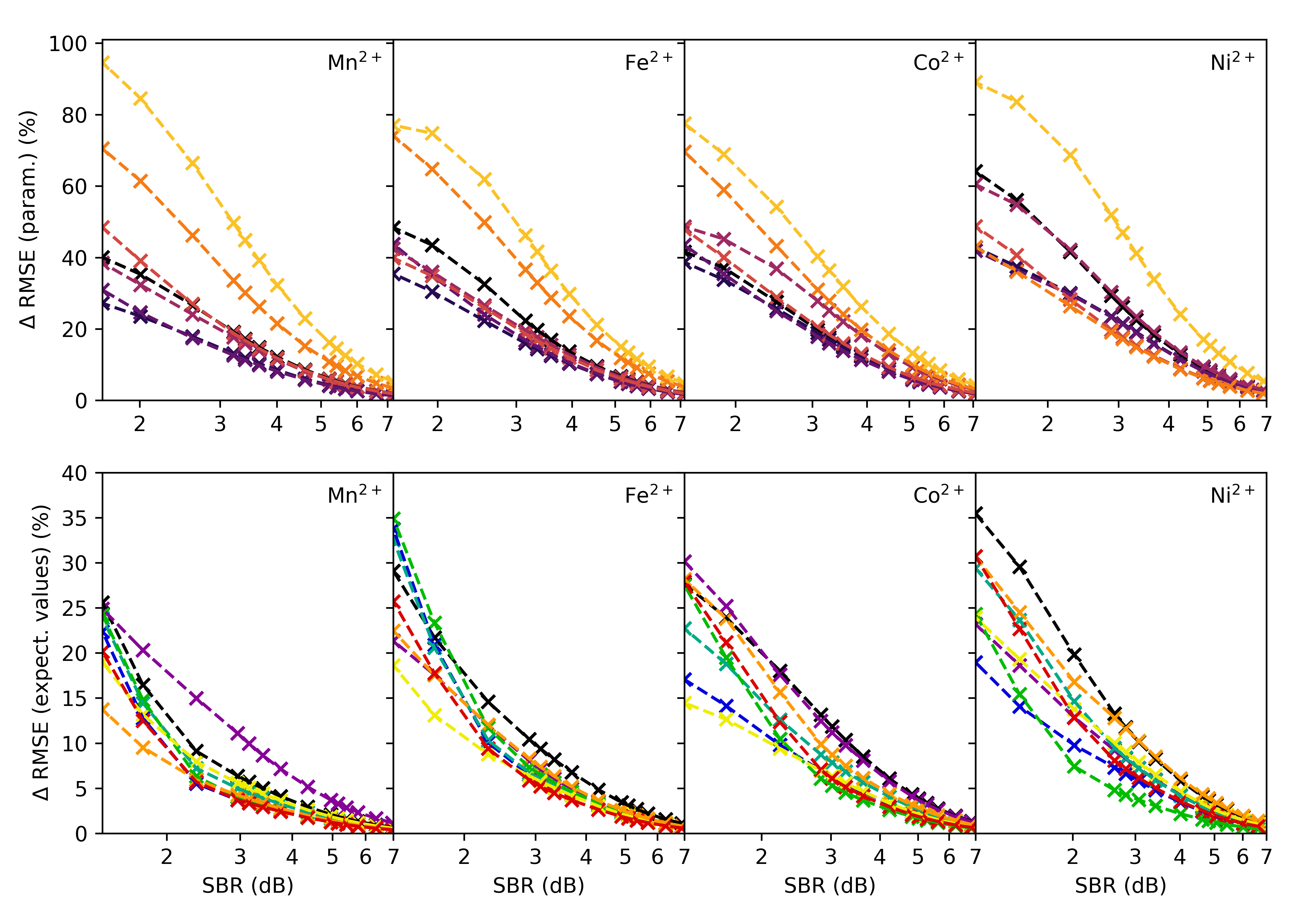}
\caption{(Color online) Influence of linear background in test spectra on the A2pX-T2EP's analytic (parameter estimation) and predictive (expectation value estimation) performance measured in $\Delta$RMSE. Same curve labeling as in Fig.~\ref{fig:rmse_noise} applies. }
\label{fig:bgerr}
\end{figure}

At this point, the effect of small changes in spectra, which could also arise from noise and might lead to noticeable changes in the output vector, needs to be analysed. In addition, background contributions are common in X-ray spectra and various mechanisms, such as  solvents or secondary excitation processes and sometimes excitations to other edges, can contribute to them \cite{Kubin2018a,DeGroot1997}. 
For instance, in the case of Ni$^{2+}$, the intensity around 867 eV for 2$p$ XAS of NiO is ascribed to excitation of electrons from 2p$_{3/2}$  to free states \cite{Haverkort2012}, see Fig.~\ref{fig:tmo}. Besides, intensities beyond 872 eV show an approximately  constant background signal. 
Hence, random noise (or white noise) and background contributions in spectra must be considered on how they affect the capabilities of the ANN methods for practical applications. 

The influence of white noise and background contributions in spectra on the ANN's performance is tested with different background models and varying signal-to-noise ratios (SNR) and signal-to-background ratios (SBR) in  normalized test spectra. 
When adding white noise, the ANN output vectors obtained from a spectrum with noise (y$_p^{noise}$) are compared to the output vectors obtained for the same spectra without noise. The resulting RMSEs are reported. To distinguish these RMSEs  from the performance errors of the ANNs discussed above, noise-based errors, that evaluate residuals $y_p^{noise}-y_p$, are labeled $\Delta$RMSE. 
The latter errors were averaged over 100 trials, each having 10000 test spectra.  The SNR ratio is reported as $20 \log{[(RMS(I_{XAS}) / RMS(I_{noise})]}$ where RMS is the root mean square of the 2$p$ XAS spectrum's  intensity $I_{XAS}$ or the noise contribution $I_{noise}$ to it. Background intensities were modeled with a linear function. In addition, step backgrounds were modeled by two hyperbolic tangent functions, i.e. $0.5 \beta(\tanh(\alpha(E-E_{L_i}))+1)$, a so-called soft-step function, centered at the L$_3$ or L$_2$ peaks. The factors $\alpha$ (taken as 2 eV$^{-1}$), $\beta$ and E$_{L_i}$ influence the width, the max. height and the center of the soft-step function, respectively. The background is added to the spectra of the test datasets. Then spectra with background are passed to the trained ANNs and their performance is evaluated with respect to the test parameters. The difference between parameters obtained with background and original parameters is evaluated through  SBR, that is computed like the SNR but using the background  instead of the noise intensity. The analysis is also based on $\Delta$RMSE. A max. noise range / background contributions between 1$\times 10^{-4}$\% and 2.0\% of the max. intensity of the reference spectra was considered. 

	It is noteworthy mentioning that the impact of noise on the accuracy of the ANNs becomes significantly stronger with a feature-wise (i.e. channel-wise) normalization instead of spectrum-wise normalization, which applies equally to Z-score scaling \cite{ErwinKreyszig1979} and the here used box normalization. This is caused by large noise-induced fluctuations in low-intensity regions in spectra. Still, the feature-wise normalization can generally result in better performance of the ANNs; however, their noise sensitivity may hinder or challenge real-world application.

Figure~\ref{fig:rmse_noise} shows how random noise in test spectra affects  A2pX-T2EP's capability and accuracy to determine features  of the output vector  for Co$^{2+}$, Ni$^{2+}$, Fe$^{2+}$ and Mn$^{2+}$. Data points show the mean values obtained through evaluating 100 trails each having 10000 test spectra  and error bars give their ranges (i.e. min. to max. error). The error bars are negligible  for parameter estimation while they are noticeable for expectation value prediction, especially for $<J^2>$ .
 
The effect of noise is most noticeable in Mn$^{2+}$   while its impact decreases slightly for parameter estimation and significantly for  expectation value estimation with increasing Z number. 
Substantial uncertainties are in $\Gamma_2$ followed by  $\Gamma_1$ for all tested ions.   For SNR larger than ca. 4 dB, this correlates to some degree with the overall feature resolved accuracy of the ANNs, given in Tab.~\ref{tab:res03}. 
Estimating convolution factors,  D$_q$ and  screening factors is also sensitive to random noise. For instance, for the Ni ion, the $\Delta$RMSEs of the Ballhausen parameters and S$_1$ increase faster with noise than $\Gamma_1$. 
In contrast, expectation value prediction is less affected by random noise. At low SNR, the $\Delta$RMSEs are by a factor of ca. 4 to 10 smaller compared to the ones of the parameter determination. The largest $\Delta$RMSEs are seen in $<J^2>$ for all ions, while $<S^2>$, $<L^2>$ and orbital occupations are less affected among expectation values. 
 The observed $\Delta$RMSEs are smaller than the RMSEs at large SNR, which should not result in severe limitation for practical applications as long as the SNR is reasonable. 
  
Figure \ref{fig:bgerr}  compares the effect of SBR probed by linear background contributions     on the $\Delta$RMSEs for all tested ions in D$_{4h}$ CFs. The max. $\Delta$RMSEs are much larger (by a factor of 15 for parameter and 75 for expectation value estimation at ca. 3 dB) than for  random noise. Therefore the magnitude of the $\Delta$RMSEs indicates sever  limitations at small SBR.  At large SBR, the $\Delta$RMSEs decrease to an acceptable size. $\Gamma_2$ shows the largest dependence, followed by $\Gamma_1$ and the screening factors. Only for Ni$^{2+}$, the $\Delta$RMSE of $\Gamma_1$ is smaller than some RMSE values of the ligand-field and atomic parameters. 
The results for the step function backgrounds are given in SI in Fig.~\ref{fig:stepbackground}. The trends follow roughly the same as for the linear background. Two points are noticeable that are: i) the magnitude of $\Delta$RMSE caused by background signals is much larger than for white noise, and ii) there is a weak correlation between feature accuracy in the testdata and an increase of $\Delta$RMSE with an increasing background. Moreover, step backgrounds centered either at L$_3$ or L$_2$ lead to comparable values of $\Delta$RMSE. Hence, accurate background reduction can be of equal importance for L$_3$ and L$_2$ features with small SBR.

It can be concluded that removal of background contributions from experimental spectra is essential to increase the usefulness of the ANN-based method in analyzing  XAS spectra at the $L$-edge. In contrast, noise removal becomes only necessary for very noisy signals and for high accuracy predictions for some expectation values, while spin and orbital momenta as well as orbital occupations are not significantly affected by noise. 

 \subsection{Revisiting the 2$p$ XAS of TM compounds with ANN}
 
\begin{figure}[!htbp]
\centering
\includegraphics[width=1.0\textwidth]{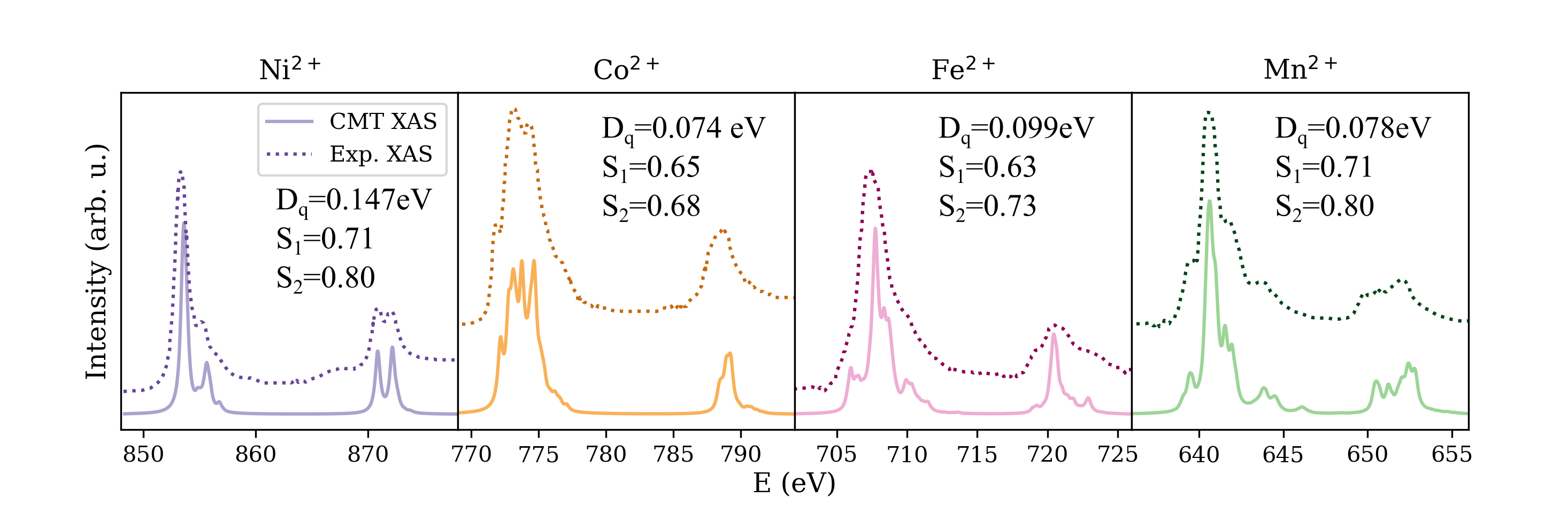}
\caption{(Color online) Comparison of experimental $L$-edge XAS spectra (dotted) of NiO \cite{Alders1998}, CoO \cite{Groot1993}, FeO \cite{Regan2001} and MnO \cite{Gilbert2003} and the simulated spectra (solid) based on the analysis of the trained six HLs medium width ANN. }
\label{fig:tmo}
\end{figure}

Furthermore, the ANNs are tested on experimental spectra. Recomputed spectra served to validate the results. The theoretical and experimental spectra were shifted in energy to facilitate comparison.
For A2pX, the results for TM oxides are presented in Fig.~\ref{fig:tmo}.
Experimental spectra are taken from Ref. \cite{Alders1998,Groot1993,Regan2001,Gilbert2003} for NiO, FeO, CoO and MnO.
 A2pX determined $10D_q$ values for Ni$^{2+}$, Fe$^{2+}$, Co$^{2+}$ and Mn$^{2+}$ ions are 1.47, 0.99, 0.74 and 0.78 eV, respectively.  The values are in reasonable agreement with previous results \cite{Luder2017a,Haverkort2012}.  The determined screening factors are between 0.63 and 0.8 (details are given in Fig.~\ref{fig:tmo}). 
 The ANN  yielded the exchange screening (S$_2$) being less than 
the Coulomb screening (i.e. S$_2$  $>$ S$_1$).
The agreement of the computed XAS spectra  with the experimental spectra shown in  Fig.~\ref{fig:tmo} is remarkable, considering the simplicity of the underlying methods. 
Even small features, for instance, at 780 eV in Co$^{2+}$, 710 eV in Fe$^{2+}$ and 645 eV in Mn$^{2+}$ are qualitatively reproduced. In addition, the L$_3$-L$_2$ branching ratios are comparable, which  confirms the ANNs accuracy. Limitations are present mostly in the peak widths caused by underestimated broadening and neglecting  effects of temperature.

\begin{figure}[!htbp]
\centering
\includegraphics[width=0.85\textwidth]{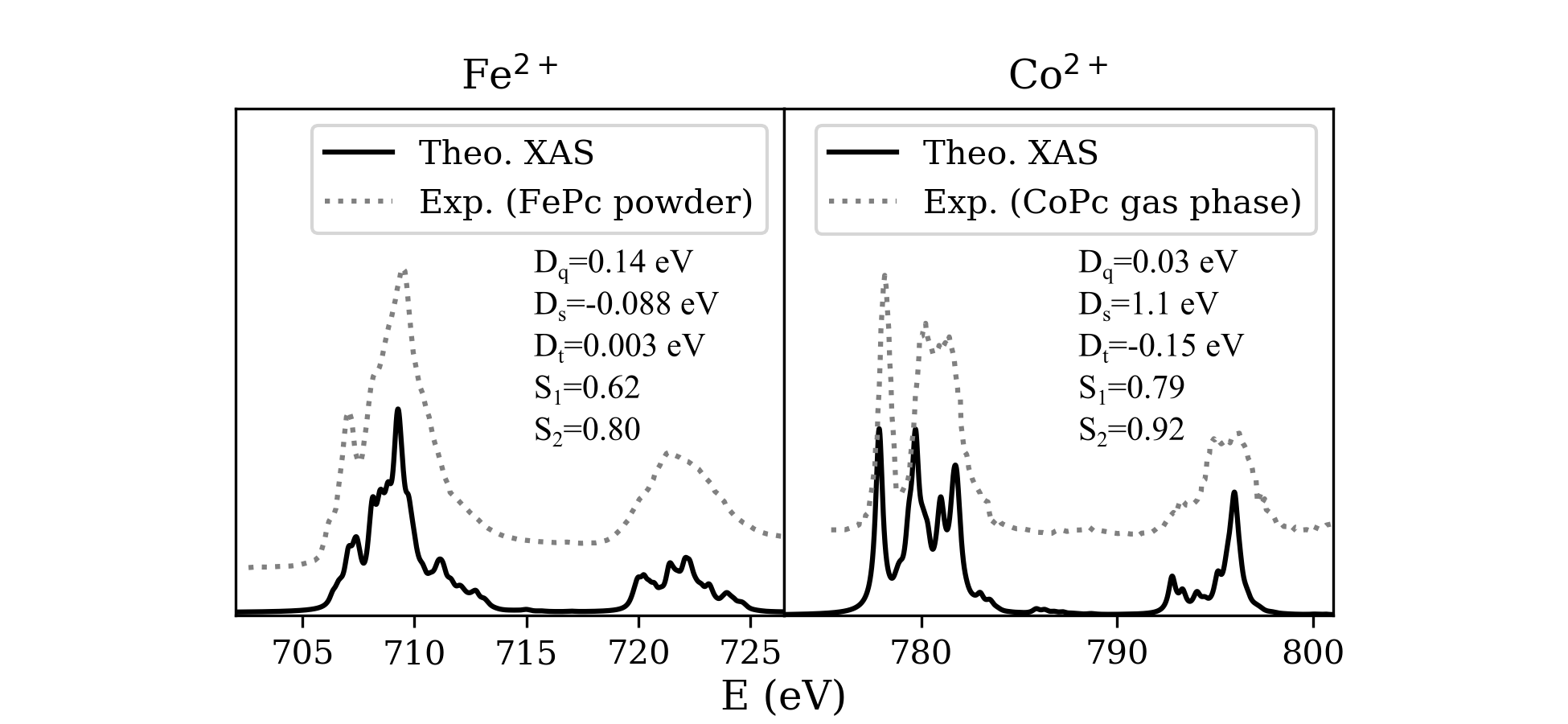}
\caption{(Color online)   Comparison of experimental $L$-edge XAS spectra (dotted) of FePc (powder) \cite{Miedema2009} and CoPc (gas phase) \cite{Zhang2017} with the simulated spectra (solid) that are based on the evaluation using  the six HLs medium width ANN.  }
\label{fig:mpc}
\end{figure}

More challenging is the $L$-edge XAS of TM ions with D$_{4h}$ symmetric CF like the ones in Co- and Fe-phthalocyanine (Pc), for which experimental spectra are taken from Ref. \cite{Zhang2017} and \cite{Miedema2009}, respectively.  
A2pX determined parameters of the model Hamiltonian are given in Fig.~\ref{fig:mpc} for the $L$-edge XAS of FePc and CoPc. The figure  compares the theoretical (recomputed) with the experimental spectra.

For FePc, the agreement between  experimental and computed spectrum  is acceptable, but the same limitations (fixed temperature and underestimated broadening) as in the case of simple TM oxides seem to apply. The computed spectrum is based on the ANN analysis of the experimental spectrum. The similarity between the spectra is seen in, e.g., the tail of the 
L$_3$ peak,  a small feature at 706.5 eV and the shape of the main peak. Also, the  L$_2$ peaks   have an overall similar spectral shape. The electronic configuration of the $d$-subspace determined by  A2pX is  $a_{1g}^1b_{1g}^1b_{2g}^2e_{g}^2$ is a  singlet state. 
It should be noted that this result is based on CMT and presents the best fit to the experimental spectrum that the ANN can produce. Previous works determined  similar  but also different electronic configurations in FePc \cite{Coppens1983,Kroll2012,Miedema2009}, for which several authors have discussed the true ground state of FePc  in recent years. At this point, the ANNs cannot exclude other electronic configurations of FePc since the results are limited to CMT.

To accurately reproduce the spectrum of  CoPc, the parameter ranges and the size of the dataset needed to be increased. A dataset size of more than 100000 spectra with ranges for $D_q=[0.0,0.3]$ eV, $D_s=[-2.0,2.0] $ eV and $D_t=[-1.0,1.0]$ eV were used. This gives a similar per-feature dataset size of ca. 15000 as the A2pX-T2EP case and resultsed in equivalent performance of the ANN as discussed above. Then, the computed CoPc spectrum belonging to a doublet  ground state ($a_{1g}^1b_{1g}^1b_{2g}^1e_{g}^4$) is reproduced and agrees with previously  reported results \cite{Zhang2017}.

 \begin{figure}[!htbp]
\centering
\includegraphics[width=1.0\textwidth]{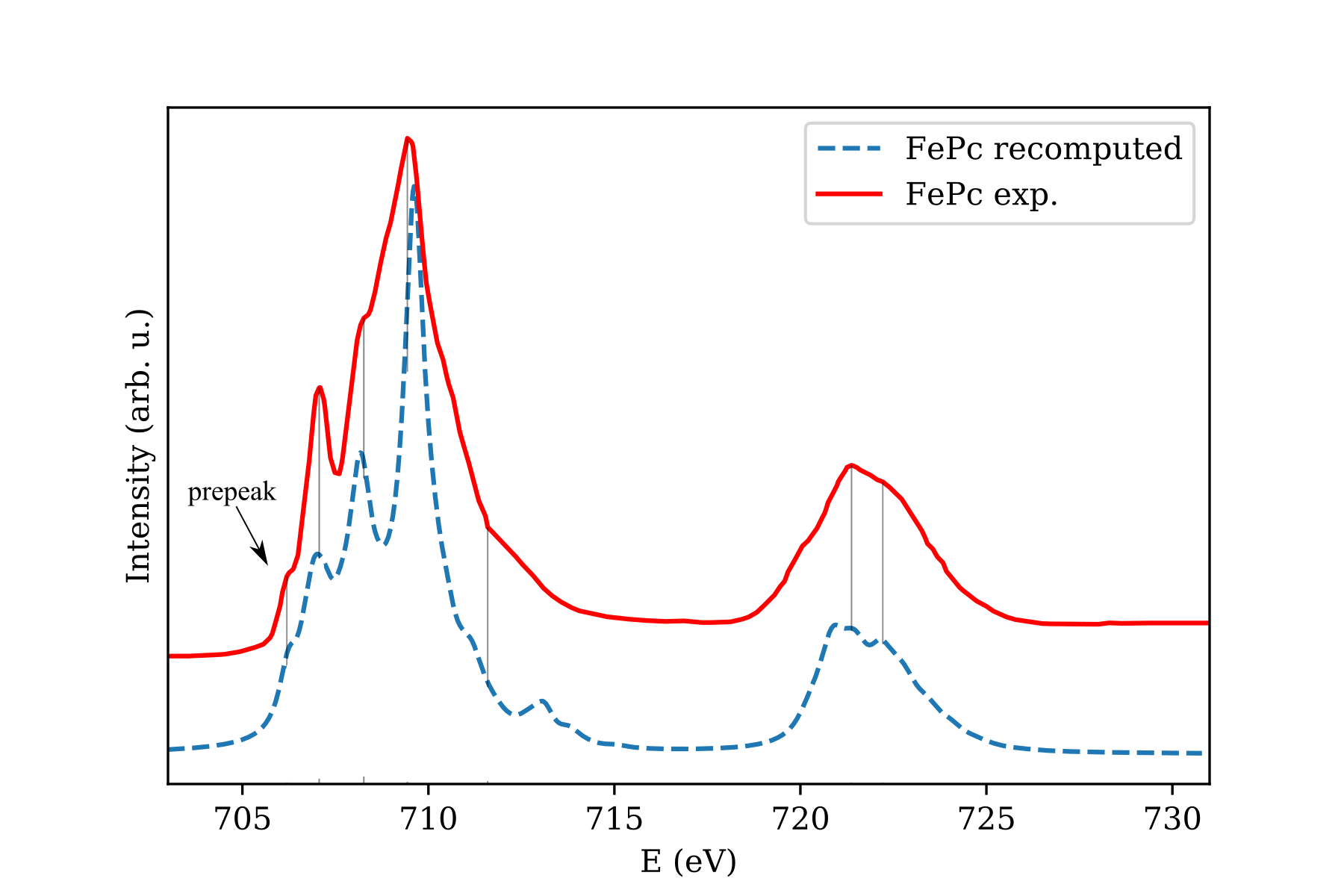}
\caption{(Color online) Comparison of experimental (red) \cite{Miedema2009} and recomputed spectra (dashed blue) based on the results of A2pX-T2 for FePc. Vertical lines are given for reference of spectral feature positions.}
\label{fig:fepc_v4}
\end{figure}

The recomputed spectra based on the analysis given by A2pX display a few features that can be improved, such as sharp peaks of high intensity in L$_2$ and an overall rougher appearance due to underestimated broadening effects. 
A2pX-T2 is tested on the already introduced experimental 2$p$ XAS spectra of Co$^{2+}$ and Fe$^{2+}$ ions. As indicated by the above results, A2pX-T2 will yield similarly accurate outcomes as A2pX-T2EP. 
Moreover, 2$p$ XAS spectra of Co$^{2+}$ at the $L$-edge are known to display noticeable temperature dependence \cite{DeGroot1997,Csiszar2005}. Applying A2pX-T2 should improve on these issues. Still, it must be understood that background contributions can severely affect the outcome.  The discussion is based on resemblance between recomputed spectra and experimental spectra. Only 2p-to-3d transitions that contribute to the experimental spectrum, as described by Eq.~(\ref{eq:spect}), must be present for optimal performance of the ANNs. Therefore, other contributions such as background intensities have to be removed.  Still, the description itself is incomplete given, for instance, by missing charge transfer  (CT) effects (see above discussion). Overcompensation of the ANNs for missing physical effects (e.g. the earlier mentioned CT) may occur. 

For background subtraction in the 2$p$ XAS spectrum of FePc, one step function was centered at L$_3$ at ca. 710 eV and the height ($\beta$) was chosen to approx. level the intensity before L$_2$ at 0 and the other function was center at L$_2$ at around 722.3 eV with a height that levels the intensity beyond 728 eV at 0. No noise reduction was applied. The temperature was set to 8 K, at which the experiment was performed \cite{Miedema2009}, and the experimental FWHM equaled 0.1 eV. 

A2pX-T2 determined D$_q$=0.13 eV, D$_s$=0.35 eV, D$_t$=0.037 eV, S$_1$=0.62 and S$_2$=0.86. The lifetime broadening factors are $\Gamma_1$ = 0.59 eV and $\Gamma_2$ = 0.76 eV. Compared to the A2pX results,  D$_s$ experiences a large change, while the other atomic and ligand field parameters are similar to the ones determined by A2pX.  In an ealier study, Miedema \textit{et al.} performed the analysis of 2$p$ XAS of FePc using planar symmetry that relates values of D$_s$ and D$_t$ to the value of D$_q$. Concerning the formal rule for planar symmetry and $D_q$, the A2pX-T2 determined values are mismatched, D$_s$ by a factor of 2 and D$_t$ by a factor of 0.5. Note that  chemical environments can lead to a reduction of symmetry in FePc. Previously, Snezhkova \textit{et at.} detected symmetry breaking  in FePc through interaction of the molecule with its chemical surrounding, i.e. when adsorbed on Cu(111) surface, by XPS measurements and first-principle calculations \cite{Snezhkova2015}. While a mismatch of parameters, especially for small values, cannot be excluded, a reduced symmetry may at least partly contribute to the yielded A2pX-T2 outcome. 

 Figure~\ref{fig:fepc_v4} shows the recomputed 2$p$ XAS spectrum  and the experimental spectrum of FePc. The agreement between recomputed and experimental spectra is satisfactory, which demonstrates that A2pX-T2 performed well  for this case. Compared to the recomputed spectrum employing the output vector of A2pX, the agreement in the spectral shape of L$_2$  improved. The A2pX-T2-based recomputed spectrum reproduces the steep increase between 720 to 722 eV followed by a gradual decrease between 722 and 725 eV. The similarity in L$_3$ did somewhat increase. The pre-peak at around 707 eV, the slight shoulder at ca. 712.5 eV and the main peak at 710 eV (although the latter peak is slightly shifted to higher energies and overestimated  in intensity) are still present. A peak at 711 eV is overestimated by a bit. Moreover, a small peak at 713 eV appeared. 

  For FeO, the temperature was set to 848 K, a temperature at which FeO is a stable phase, and an experimental FWHM of 0.1 eV is assumed. Note that there was little difference in the recomputed spectrum for input vectors with temperatures at 400 K and 848 K passed to the ANN. In contrast, the results show larger differences if low temperatures, e.g. 10K, are used.
  A subtraction of the stepped background was applied with two soft-step functions. One step function was centered at L$_3$ at ca. 713 eV and the height ($\beta$) was chosen to slightly lower intensity before L$_2$ and the other function was center at L$_2$ at around 725.3 eV with a height that levels the intensity beyond 730 eV closer to 0. Also, the signal was truncated (i.e. set to 0) before 706 eV. No noise reduction was applied. 

D$_q$ was determined as 0.022 eV, the screening parameters as $S_1= 0.90$ and $S_1= 0.92$, and the lifetime broadening factors are 0.77 ($\Gamma_1$) and 1.03 eV ($\Gamma_2$'). The determined value of D$_q$ for TMO is underestimated with regard to previous reports \cite{DeGroot1997} and the A2pX results, which usually give values around 0.1 eV (or 1 eV for 10D$_q$). It is well-known that there can be a certain insensitivity towards D$_q$ values in 2$p$ XAS spectra, especially for TMOs,  allowing a range of values to reproduce experimental results. Here, the experimental and recomputed spectra are similar. This potentially points towards a more delicate relation between ligand field parameters, screening factors and lifetime broadening (potentially also for D$_{4h}$ systems) in the underlying MHA as commonly assumed by fixed values of screening factors of 0.8 or restrictions such as S$_1$=S$_2$.    

\begin{figure}[!htbp]
\centering
\includegraphics[width=1.0\textwidth]{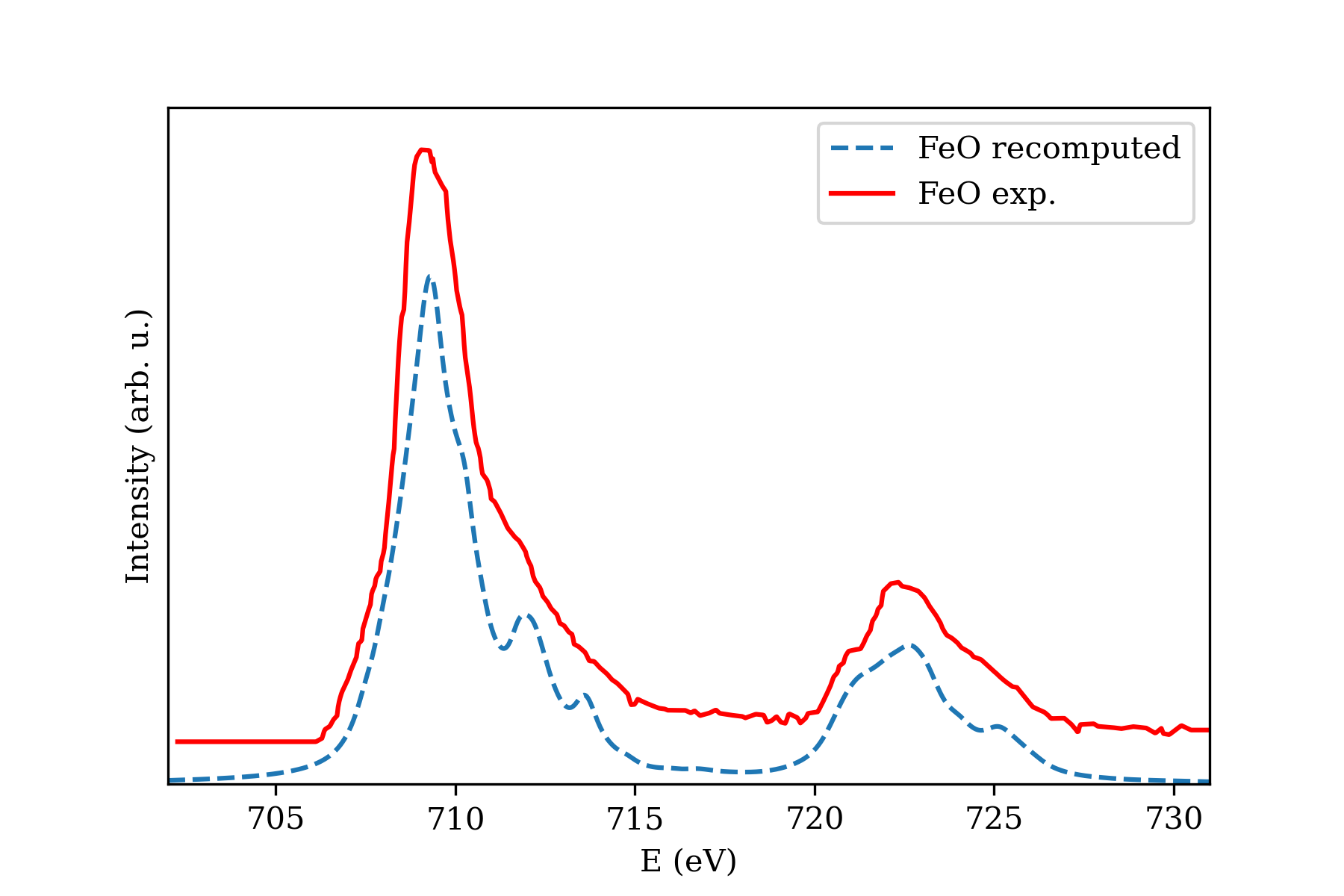}
\caption{(Color online) Comparison of experimental (red)  \cite{Regan2001} and recomputed spectra (dashed blue) based on the results of A2pX-T2 for FeO. }
\label{fig:feo_v4}
\end{figure}

Figure~\ref{fig:feo_v4} shows the comparison between experimental and recomputed (based on A2pX-T2 results) 2$p$ XAS spectra of FeO. There is a noticeable improvement for both L$_3$ and L$_2$ peaks in the recomputed 2$p$ XAS spectrum compared to the results in  Fig.~\ref{fig:tmo}. The overestimation of the main peaks in L$_3$ and L$_2$ was reduced. Similarly, the pre-slope starting at 706 eV and the width of the main peak are well-matched in L$_3$. The shoulders at ca. 711 and 714 eV are somewhat over-pronounced but follow  the tail of L$_3$ in their intensity ratios. Especially the latter peak at 714 eV is sensitive to the value of D$_q$. At larger D$_q$ values, the peak would shift to higher energies which would diminish the agreement for the given parameter set.  At ca. 721 eV, the close-by main peak and the tail are well represented in L$_2$.

Background subtraction in the 2$p$ XAS spectra of CoPc and CoO was performed with soft step-functions to align the onsets of the L$_3$ and L$_2$ peaks at 0 and to remove background contributions beyond L$_2$ transition intensities.
For CoPc, A2pX-T2 determined the parameters  D$_q$=0.0043 eV, D$_s$=1.00 eV, D$_t$=-0.13 eV, S$_1$=0.91 S$_2$=0.79, $\Gamma_1$=0.44 eV and $\Gamma_2$'=0.75 eV at a FWHM of 0.2 eV and the temperature set to 708 K at which the experiment was performed.  
For CoO, A2pX-T2 determined the parameters as  D$_q$=0.0083 eV, S$_1$=0.82 S$_2$=0.81, $\Gamma_1$=0.68 eV and $\Gamma_2$=0.73 eV at a set FWHM of 0.2 eV and a set temperature of 303 K.

\begin{figure}[!htbp]
\centering
\includegraphics[width=1.0\textwidth]{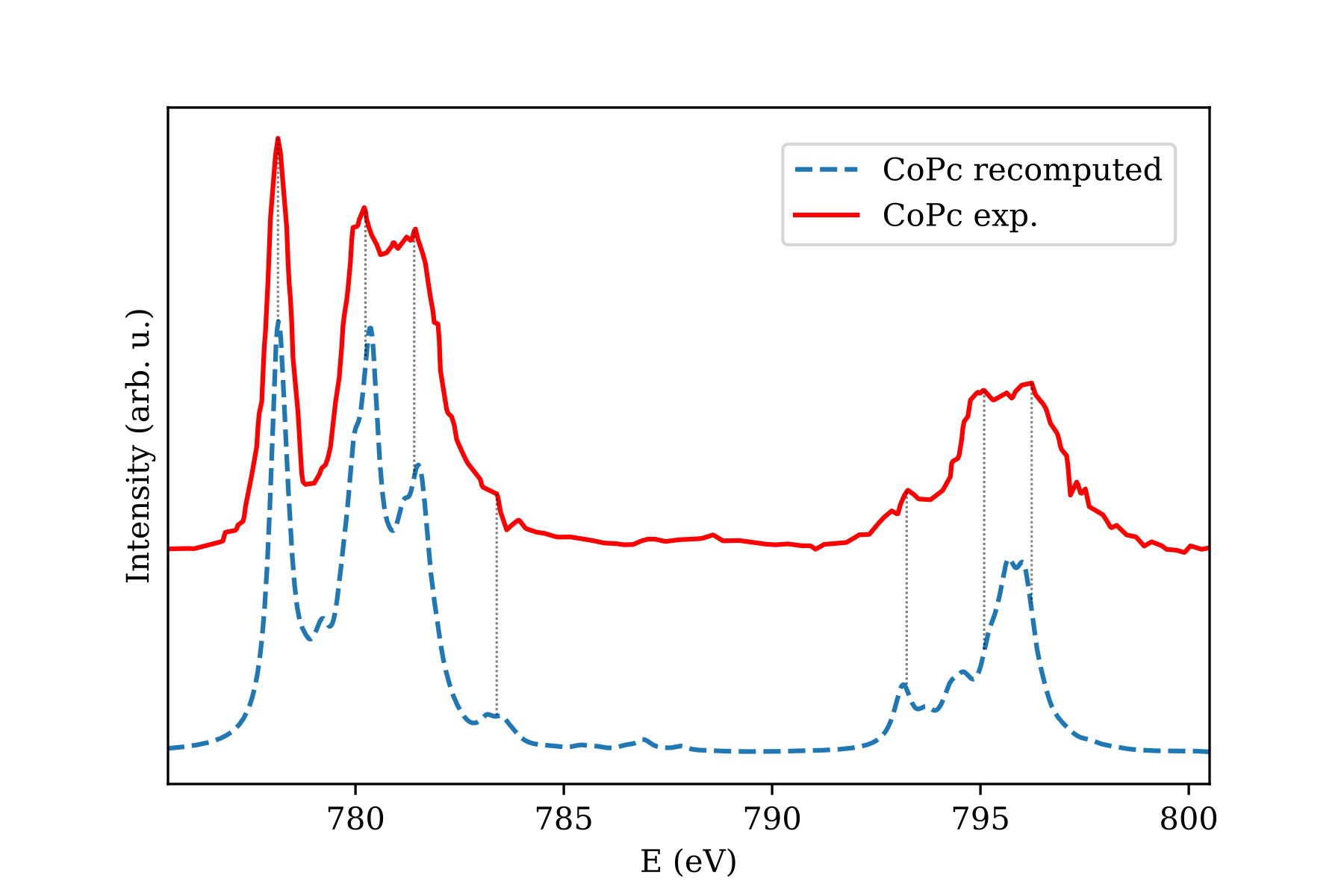}
\caption{(Color online) Comparison of experimental (red) \cite{Zhang2017} and recomputed spectra (dashed blue) based on the results of A2pX-T2 for CoPc. Vertical lines are given for reference of spectral feature positions. }
\label{fig:copc_v4}
\end{figure}

\begin{figure}[!htbp]
\centering
\includegraphics[width=1.0\textwidth]{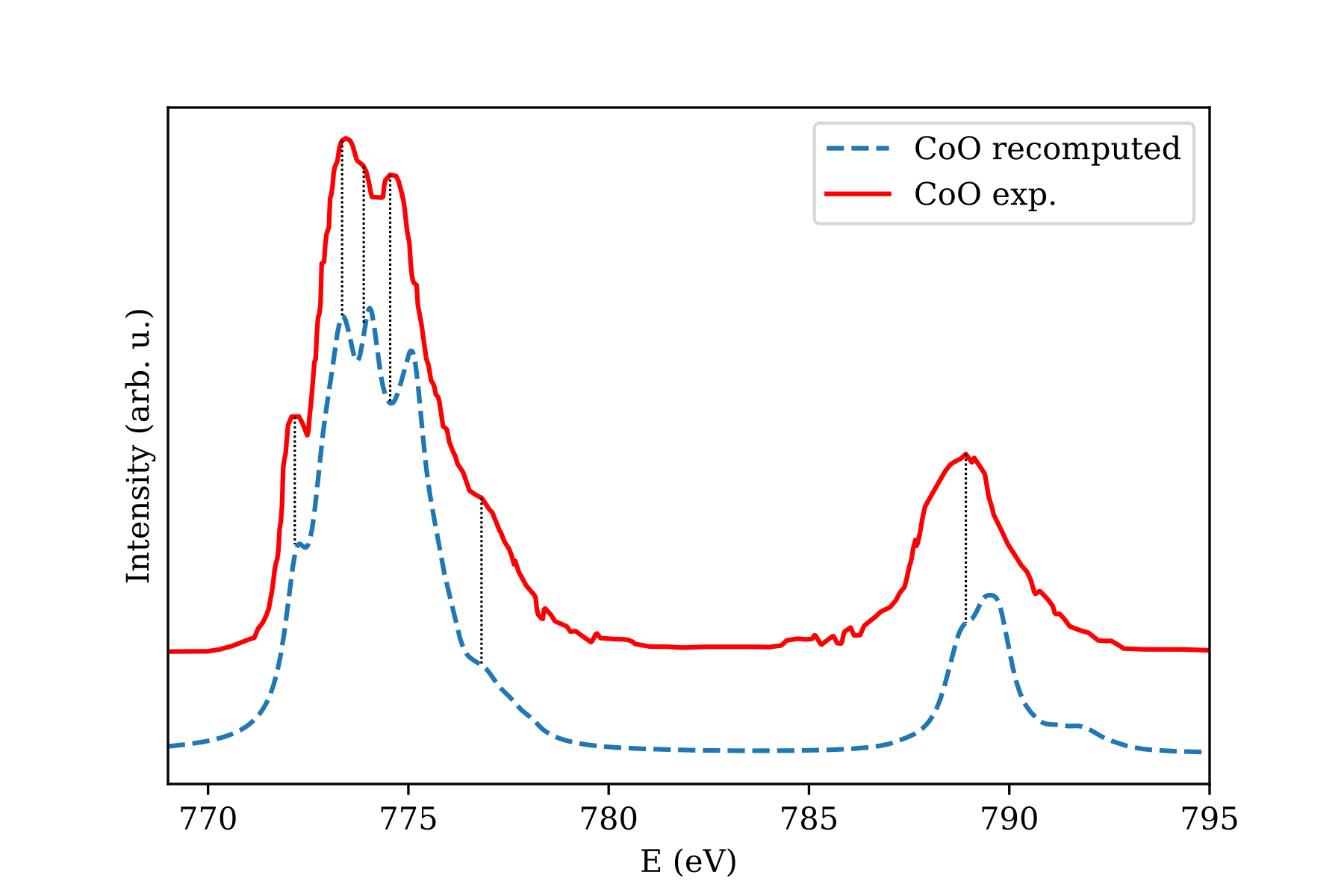}
\caption{(Color online) Comparison of experimental (red)  \cite{Groot1993} and recomputed spectra (dashed blue) based on the results of A2pX-T2 for CoO. Vertical lines are given for reference of spectral feature positions. }
\label{fig:coo_v4}
\end{figure}

The recomputed spectrum of CoPc based on the A2pX-T2 analysis is given in Fig.~\ref{fig:copc_v4}. The enhanced agreement to the experimental gas-phase spectrum, compared to the one given in Fig.~\ref{fig:mpc}, appears to be mostly due to an increased broadening of transition intensities and increased temperature. Other minor changes in intensity distribution result from changes in other parameters. The first peak at 778.2 eV and the following two peaks still match well in their max. intensity and energy spacing. Even a shoulder at 783.5 eV is reproduced, while L2 features have less agreement, in particular, at 793.5 eV and the peak structure between 795 and 797 eV. Figure~\ref{fig:coo_v4} shows the recomputed spectrum of CoO, which also displays a strong resemblance to the experimental spectrum. The main features of L3 and L2 are captured, including smaller shoulders at ca. 772.5, 777 and 792 eV.

Although the agreement between the experimental and simulated spectra of the selected test cases is encouraging, the limits of the applied method should be communicated. There is a dependence on the database and its parameter ranges, e.g., the A2pX case of CoPc, reflecting the well-known extrapolation weakness of ANNs. Another factor is the underlying theory used to construct the database. The presented ANN approach is based on CMT and the ANNs can only gain insights that any CMT-based XAS method would have produced, in particular, charge transfer effects are neglected, which may cause the ANN to compensate for this effect.
 
While the extensions of A2pX, including temperature and FWHM, increase the similarity between experimental and computed spectra, agreement of the MHA parameterization with those of previously published results is somewhat limited through potential influence of, for instance, the lack of charge transfer effects in the approach or the alignment between the experimental and computed spectra.  
At the same time, ANNs still demonstrate tremendous capabilities and  appear to be promising tools to determine the parameterization of a 2$p$ XAS model Hamiltonian and potentially its expectation values of angular and spin-momenta as well as d-orbital occupation from spectra without or little noise and background contributions.  The here presented analysis and application to experimental 2$p$ XAS spectra showed that  practical applications, however, can be limited by noise and especially background intensities in spectra, at least for dense ANN and the chosen data processing strategy. This is neither surprising nor unexpected as post-processing (e.g. background  and noise reduction) is still a routine part of the evaluation process for experimental 2$p$ XAS.  

The proposed ANN-based approach points out the need for improvements and continued developments that are required for routine application of (dense) ANN in 2$p$ XAS analysis; that can be an automated background subtraction that, in principle, can also be ML supported and / or a more robust approach for expectation value prediction. For these points, the normalization process and data pre-processing can be, to some extent, essential as the Z-score and box normalization indicated for noise-related errors (see above). Here,  noise reduction seems to be less significant for box normalization compared to background intensity removal.
Nonetheless, the ANN-based evaluation of experimental spectra and the consecutively performed recomputation of spectra were successfully performed. The obtained results are particularly encouraging for the metal-organic compounds with reduced symmetries, which  demonstrates the capability to accelerate analysis for a wide variety of materials in the near future. 
 Concerning $L$-edge XAS of metal-organic compounds containing TM ions, the method could provide an unbiased tool to obtain experimental electronic ground state configurations and gain insights on how different chemical environments influence them. Hence, the question if ANNs can represent the inversion function of conventional computations of 2$p$ XAS spectra of light transition metal ions is affirmed.

\section{Conclusions}
In summary, artificial neural networks  are presented that can estimate  relative energies of 3$d$-levels, screening factors of the Coulomb and exchange interaction, and core hole lifetimes directly from experimental $L$-edge spectra of  transition metal compounds.  The method is based on dense neural network architectures with six hidden layers, which work sufficiently well for various transition metal ions.  The artificial neural networks were trained and optimized on theoretical 2$p$ XAS spectra of transition metal ion employing a model Hamiltonian approach that is based on crystal-field multiplet theory. 
In general, obtained results become less accurate when more information (i.e. number of parameters) is to be extracted from a spectrum. This can partly be compensated by larger training dataset sizes and larger neural networks. In addition, temperature and an experimental convolution factor can be included as input features of the neural network to account for these effects in  spectra. Then,  simulated spectra employing the ANN results can obtain a very high degree of similarity to experimental spectra if proper background signal removal is applied. 
In combination with a 2p XAS model Hamiltonian, the gained results can be used to determine  physical properties such as the ground state electronic configuration and $d$-level occupations from experimental spectra in one-shot evaluations without the need for prior first-principles electronic structure calculations.  

\section*{Acknowledgments}
J.L. is thankful for financial support by the Ministry of Science and Technology (MOST) Taiwan under grant MOST109-2112-M-110-009 and acknowledges the National Center for High-Performance Computing for computer time and facilities.

\section*{References}

%

\newpage
\setcounter{figure}{0}
\makeatletter 
\renewcommand{\thefigure}{S\@arabic\c@figure}
\makeatother

\setcounter{table}{0}
\makeatletter 
\renewcommand{\thetable}{S\@arabic\c@table}
\makeatother

\section{Supporting Information}

The following information present further details about the performce of the developed ANNs used to evaluate 2$p$ XAS spectra and predict results of a 2$p$ XAS model Hamiltonian as presented in the main article.

\subsection{Optimizing ANNs}
\begin{figure}[!htbp]
\centering
\includegraphics[width=1.0\textwidth]{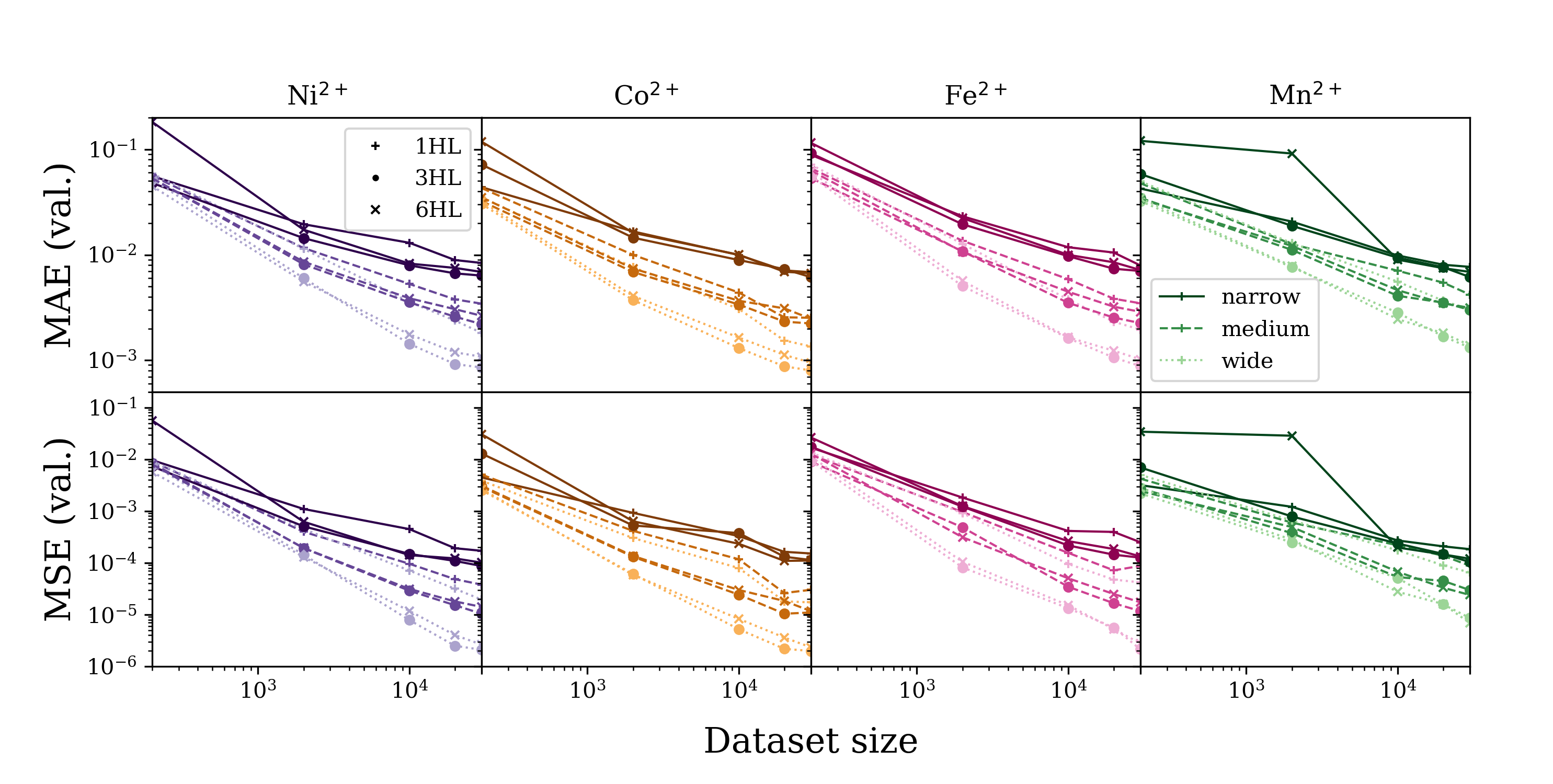}
\caption{(Color online) 
The MSE and MAE of validation data on trained ANNs with different architectures with one, three and six HLs (A2pX) for (from left to right) Ni$^{2+}$, Co$^{2+}$, Fe$^{2+}$ and Mn$^{2+}$ in O$_{h}$ and C$_{4v}$ CFs as a function of dataset size.
}
\label{fig:tre_oh}
\end{figure}

The impact of training dataset size on the MSE and MAE in validation data for O$_h$ systems is shown in Fig.~\ref{fig:tre_oh}.
Table \ref{tab:res_A2pX}, \ref{tab:res_oh_A2pXt2} and \ref{tab:res_oh_A2pXt2ep} summarize the RMSE and MAE evaluted on the test datasets and obtained with A2pX, A2pX-T2, A2pX-T2Ep, respectively, for $O_h$ systems.
  \begin{table*}[!htbp]
\caption{\label{tab:res_A2pX}  RMSE (MAE) of normalized D$_q$, S$_1$ and S$_2$, determined on the normalized test datasets (2000 spectra) for Mn$^{2+}$, Fe$^{2+}$, Co$^{2+}$ and Ni$^{2+}$ with A2pX for O$_h$ / C$_{4v}$ systems. Values are given in \%.}
\centering
\begin{tabular}{p{2.0cm}p{3cm}p{3cm}p{3cm}p{3cm}}
& Mn$^{2+}$  & Fe$^{2+}$ & Co$^{2+}$& Ni$^{2+}$ \\
\hline
\hline
D$_q$ & 0.44 (0.24) & 0.16 (0.12) & 0.16 (0.12) & 0.16 (0.12) \\
S$_1$ & 0.36 (0.24) & 0.19 (0.13) & 0.18 (0.13) & 0.20 (0.13) \\
S$_2$ & 0.30 (0.21) & 0.17 (0.12) & 0.16 (0.12) & 0.15 (0.11) \\
\hline
\hline
\end{tabular}
\end{table*}

  \begin{table*}[!htbp]
\caption{\label{tab:res_oh_A2pXt2}  RMSE (MAE) of normalized D$_q$, S$_1$, S$_2$, $\Gamma_1$ and $\Gamma_2$  determined on the normalized test datasets (10000 spectra) for Mn$^{2+}$, Fe$^{2+}$, Co$^{2+}$ and Ni$^{2+}$ with A2pX-T2 for O$_h$ and C$_{4v}$ systems. Values are given in \%.}
\centering
\begin{tabular}{p{2.0cm}p{3cm}p{3cm}p{3cm}p{3cm}}
& Mn$^{2+}$  & Fe$^{2+}$ & Co$^{2+}$& Ni$^{2+}$ \\
\hline
\hline
D$_q$ & 1.20 (0.52) & 0.88 (0.66) & 0.76 (0.47) & 0.91 (0.54)\\
S$_1$   & 0.43 (0.33)& 0.47 (0.36)  & 0.65 (0.49) & 0.77 (0.55)\\
S$_2$ & 0.40 (0.30)  & 0.41 (0.32) & 0.52 (0.40) & 0.60 (0.44)\\
 $\Gamma_1$  & 0.60 (0.43)  & 0.64 (0.49) & 0.82 (0.62) & 0.72 (0.53) \\
 $\Gamma_2$ & 2.15 (1.38) & 1.21 (0.88)   & 1.76 (1.21) & 3.00 (1.87)\\
\hline
\hline
\end{tabular}
\end{table*}

  \begin{table*}[!htbp]
\caption{\label{tab:res_oh_A2pXt2ep}  RMSE (MAE) of normalized D$_q$, S$_1$, S$_2$, $\Gamma_1$, $\Gamma_2$ and predicted the electronic configuration determined on the normalized test datasets (10000 spectra) for Mn$^{2+}$, Fe$^{2+}$, Co$^{2+}$ and Ni$^{2+}$ with A2pX-T2EP for O$_h$ and C$_{4v}$ systems. Values are given in \%.}
\centering
\begin{tabular}{p{2.5cm}p{3cm}p{3cm}p{3cm}p{3cm}}
& Mn$^{2+}$  & Fe$^{2+}$ & Co$^{2+}$& Ni$^{2+}$ \\
\hline
\hline
D$_q$ & 1.17 (0.57) &0.52 (0.30) &0.40 (0.25)  & 0.71 (0.35)\\
S$_1$  & 0.42 (0.31) & 0.76 (0.50) &  0.70 (0.52) & 0.75 (0.54)\\
S$_2$  & 0.54 (0.41) & 0.54 (0.39) & 0.54 (0.39) & 0.63 (0.45)\\
 $\Gamma_1$ & 0.88 (0.64) & 0.90 (1.53) & 0.90 (0.62) &0.76 (0.56) \\
 $\Gamma_2$ & 2.60 (1.56) & 2.63 (1.53) & 2.63 (1.53) &3.64 (2.21)\\
\hline
$<$J$^2$$>$ & 0.09 (0.01) & 1.13 (0.29) & 1.21 (0.34)& 1.26 (0.42)\\
$<$J$_z$$>$  & 2.53 (0.41) & 10.55 (2.04) & 5.76 (0.98)  & 7.81 (1.27)\\
$<$S$^2$$>$  & 0.09 (0.01) & 1.81 (0.13) & 0.81 (0.04) & 1.29 (1.06)\\
$<$L$^2$$>$  & 0.09 (0.01) & 1.79 (0.13) & 0.79 (0.07) & 0.60 (0.28)\\
n(t$_{2g}$)  & 0.09 (0.02) & 2.63 (0.55) & 3.81 (0.68) & 8.45 (1.37)\\
n(e$_{g}$)  & 0.09 (0.02) & 2.40 (0.51) & 3.64 (0.67)  & 8.08 (1.31)\\
\hline
\hline
\end{tabular}
\end{table*} 

\subsection{Analysis of spectra and prediction of electronic states}

\begin{figure}[!htbp]
\centering
\includegraphics[width=0.95\textwidth]{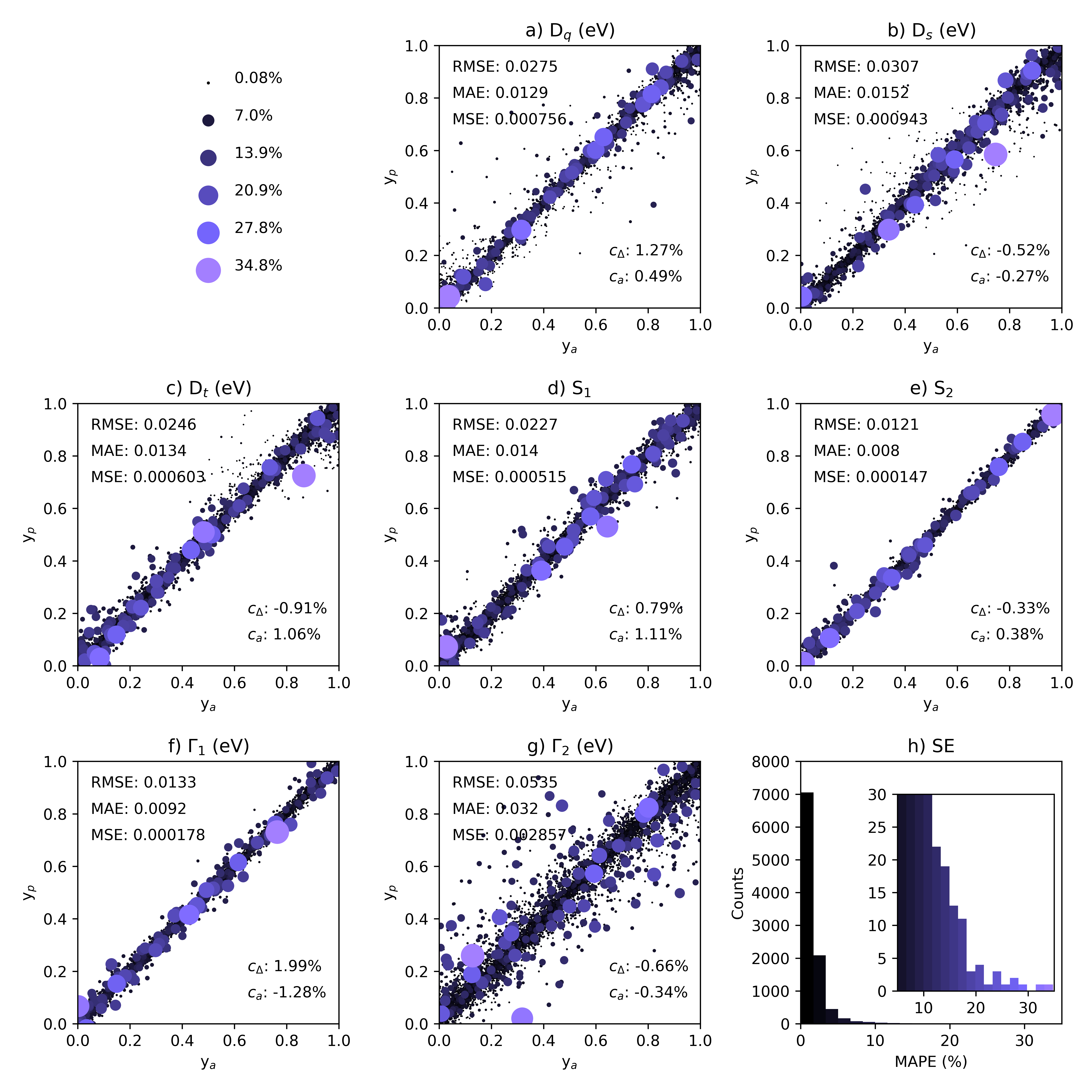}
\caption{ (Color online) From a) to g): Comparison of normalized $y_p$ and $y_a$ in D$_q$, D$_s$, D$_t$, S$_1$, S$_2$ $\Gamma_1$ and $\Gamma_2$ for  Ni$^{2+}$ with variable T and FWHM in the description vector evaluated on the test dataset (10000 spectra). Data point sizes and shades represent the SE between recomputed and reference spectra. h) The color shades are defined in the histogram of the spectrum error  (SE)  in 21 bins. The insert shows a part of the histogram with larger resolution in the counts axis.  }
\label{fig:ni_ael}
\end{figure}

\begin{figure}[!htbp]
\centering
\includegraphics[width=0.95\textwidth]{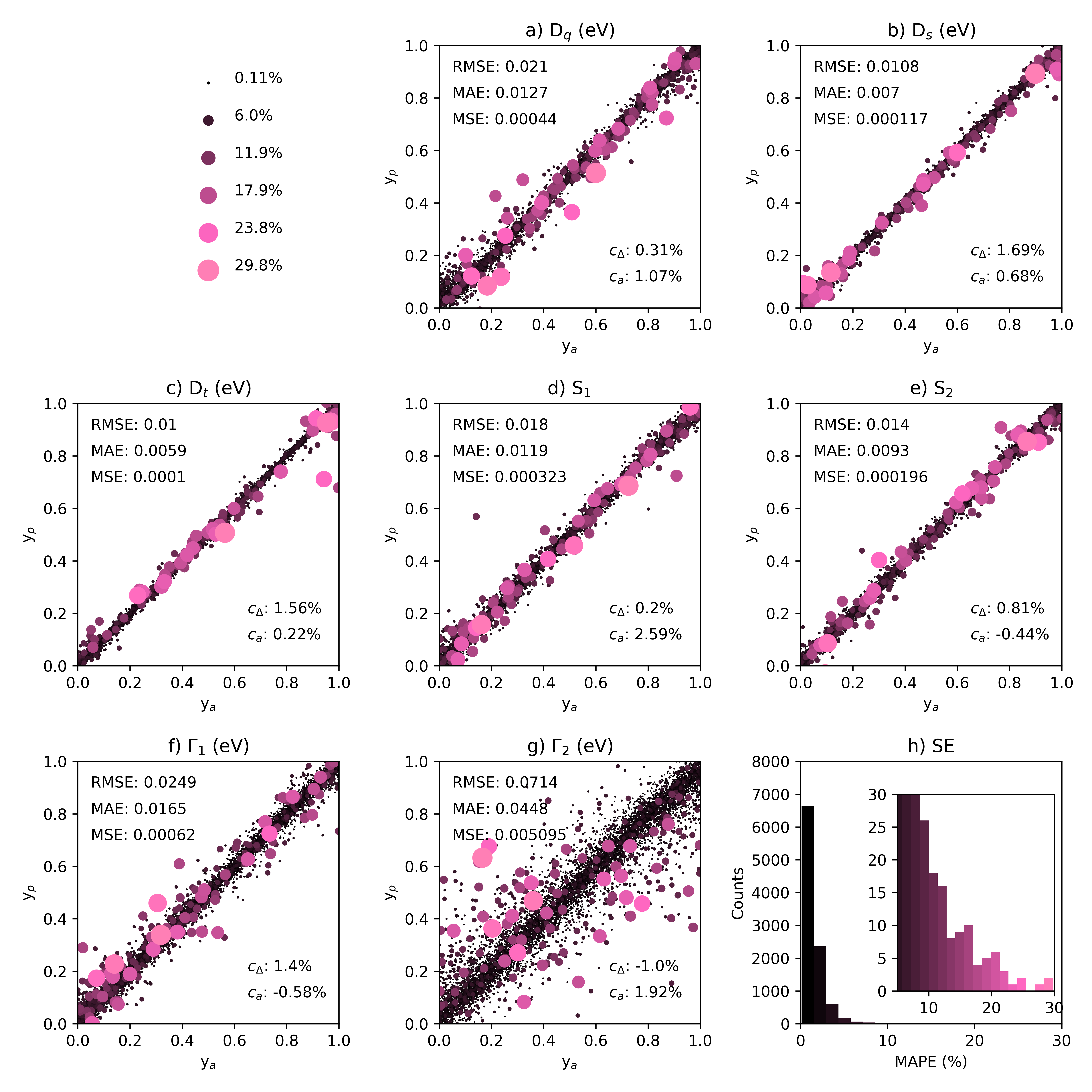}
\caption{ (Color online) From a) to g): Comparison of normalized $y_p$ and $y_a$ in D$_q$, D$_s$, D$_t$, S$_1$, S$_2$ $\Gamma_1$ and $\Gamma_2$ for  Fe$^{2+}$ with variable T and FWHM in the description vector evaluated on the test dataset (10000 spectra). Data point sizes and shades represent the SE between recomputed and reference spectra. h) The color shades are defined in the histogram of the spectrum error  (SE)  in 21 bins. The insert shows a part of the histogram with larger resolution in the counts axis. }
\label{fig:fe_ael}
\end{figure}

\begin{figure}[!htbp]
\centering
\includegraphics[width=0.95\textwidth]{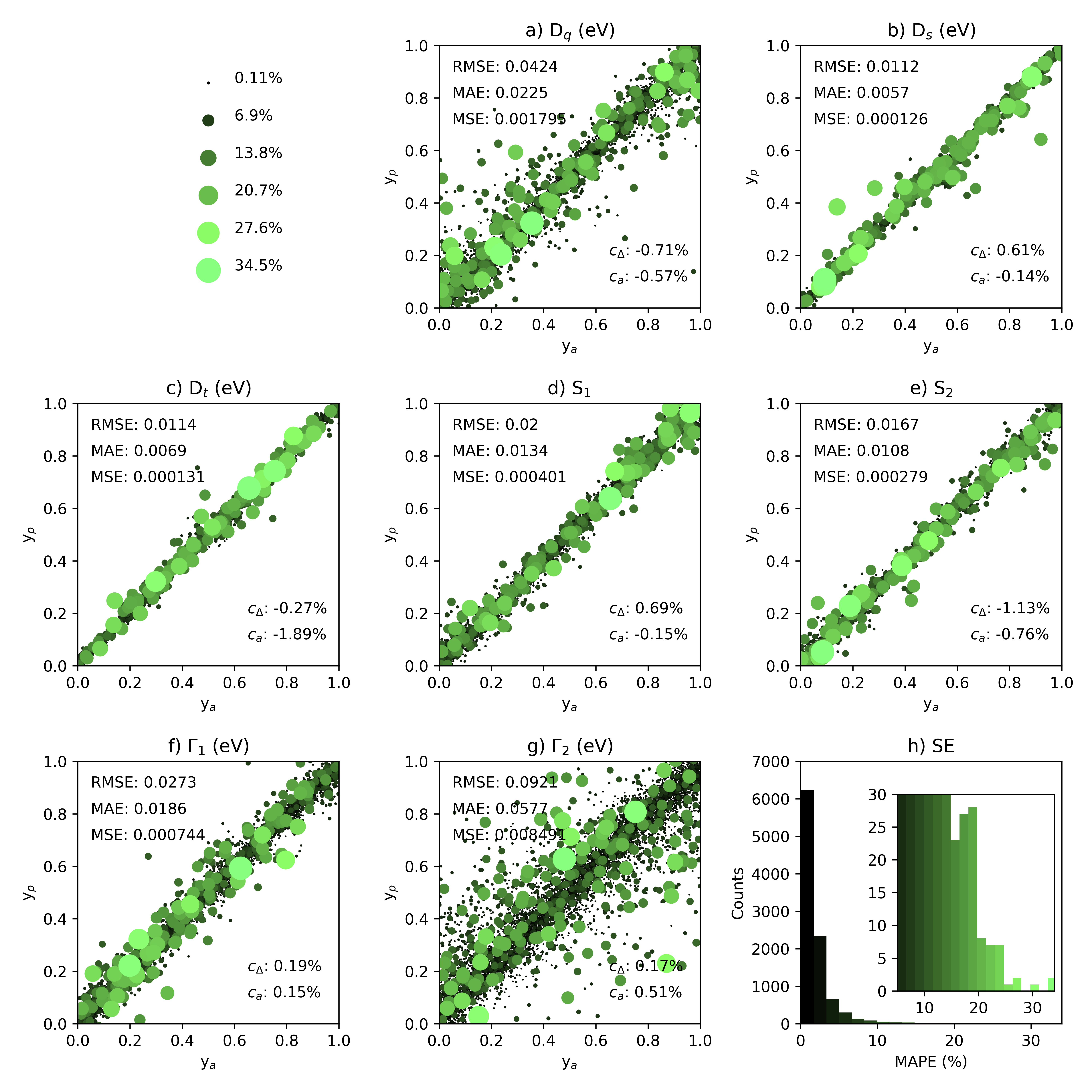}
\caption{ (Color online) From a) to g): Comparison of normalized $y_p$ and $y_a$ in D$_q$, D$_s$, D$_t$, S$_1$, S$_2$ $\Gamma_1$ and $\Gamma_2$ for  Mn$^{2+}$ with variable T and FWHM in the description vector evaluated on the test dataset (10000 spectra). Data point sizes and shades represent the SE between recomputed and reference spectra. h) The color shades are defined in the histogram of the spectrum error  (SE)  in 21 bins. The insert shows a part of the histogram with larger resolution in the counts axis. }
\label{fig:mn_ael}
\end{figure}

\begin{figure}[!htbp]
\centering
\includegraphics[width=0.95\textwidth]{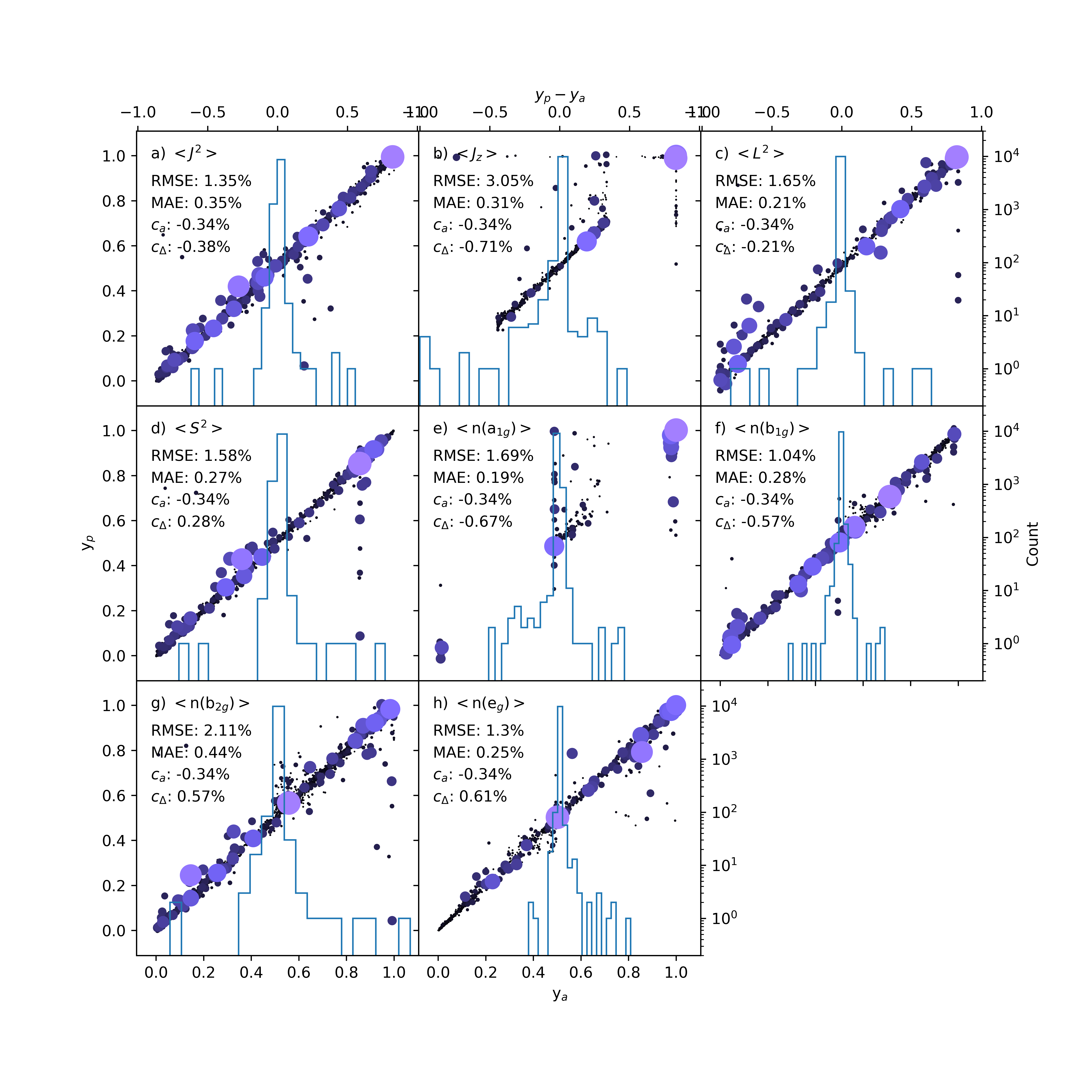}
\caption{(Color online) Comparison of normalized $y_p$ vs $y_a$ for predicted properties by A2pX-T2EP for Ni$^{2+}$ determined on test dataset (10000 spectra). The overlayed histograms show the distribution of  difference between $y_p$ vs $y_a$ on a logarithmic count scale. Pearson correlation to SE for $\Delta_y =y_p - y_a$ (c$_{\Delta}$) and y$_a$ (c$_a$). Color code and data point sizes are given in Fig.~\ref{fig:ni_ael}. The scales of the histograms are on the top right, the scales of predicted (y$_p$) vs actual data (y$_a$) are given on the bottom left.  }
\label{fig:ni_pel}
\end{figure}

\begin{figure}[!htbp]
\centering
\includegraphics[width=0.95\textwidth]{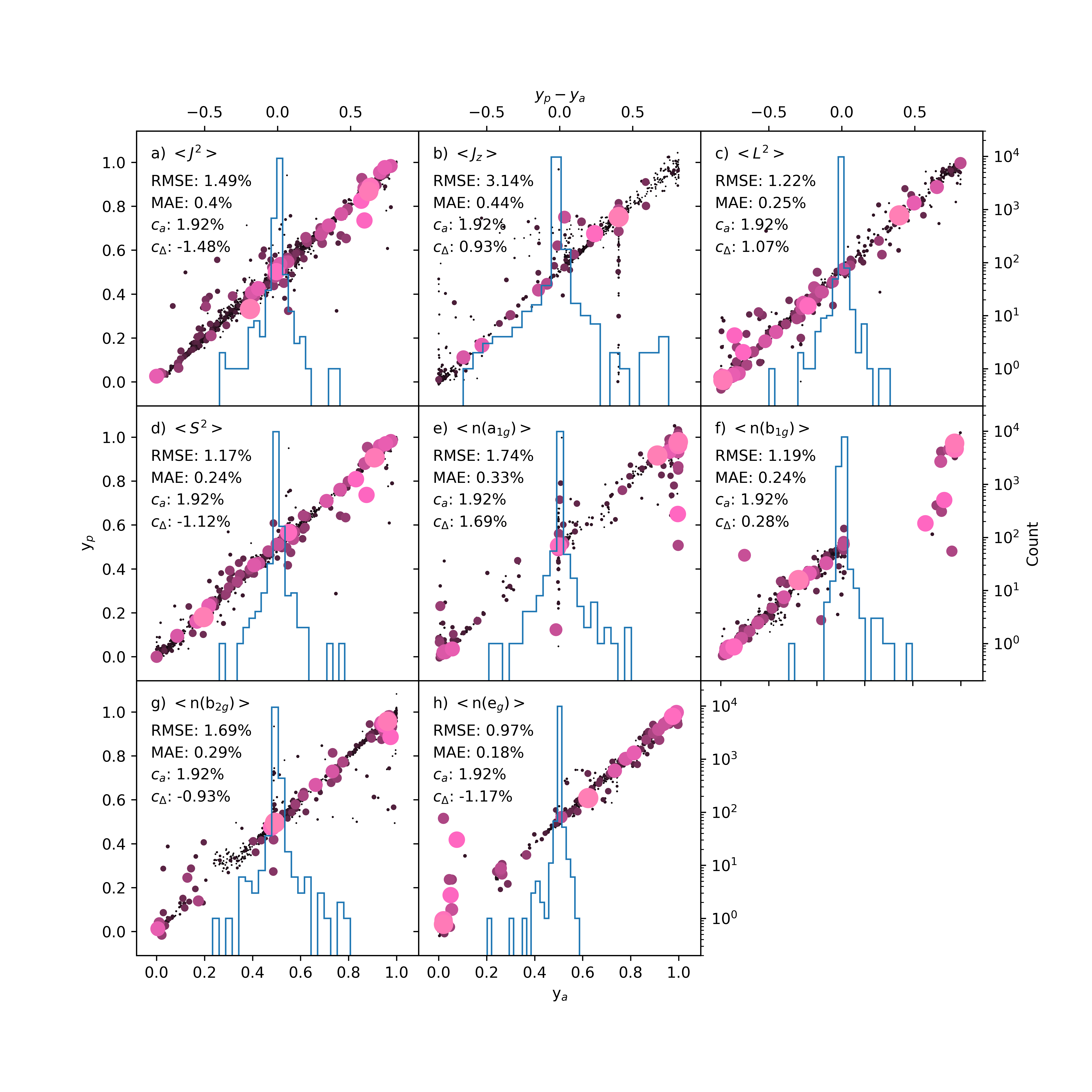}
\caption{ (Color online) Comparison of normalized $y_p$ vs $y_a$ for predicted properties by A2pX-T2EP for Fe$^{2+}$ determined on test dataset (10000 spectra). The overlayed histograms show the distribution of  difference between $y_p$ vs $y_a$ on a logarithmic count scale. Pearson correlation to SE for $\Delta_y =y_p - y_a$ (c$_{\Delta}$) and y$_a$ (c$_a$). Color code and data point sizes are as in Fig.~\ref{fig:fe_ael}. The scales of the histograms are on the right  top, the scales of predicted (y$_p$) vs actual data (y$_a$) are given on the bottom left.   }
\label{fig:fe_pel}
\end{figure}

\begin{figure}[!htbp]
\centering
\includegraphics[width=0.95\textwidth]{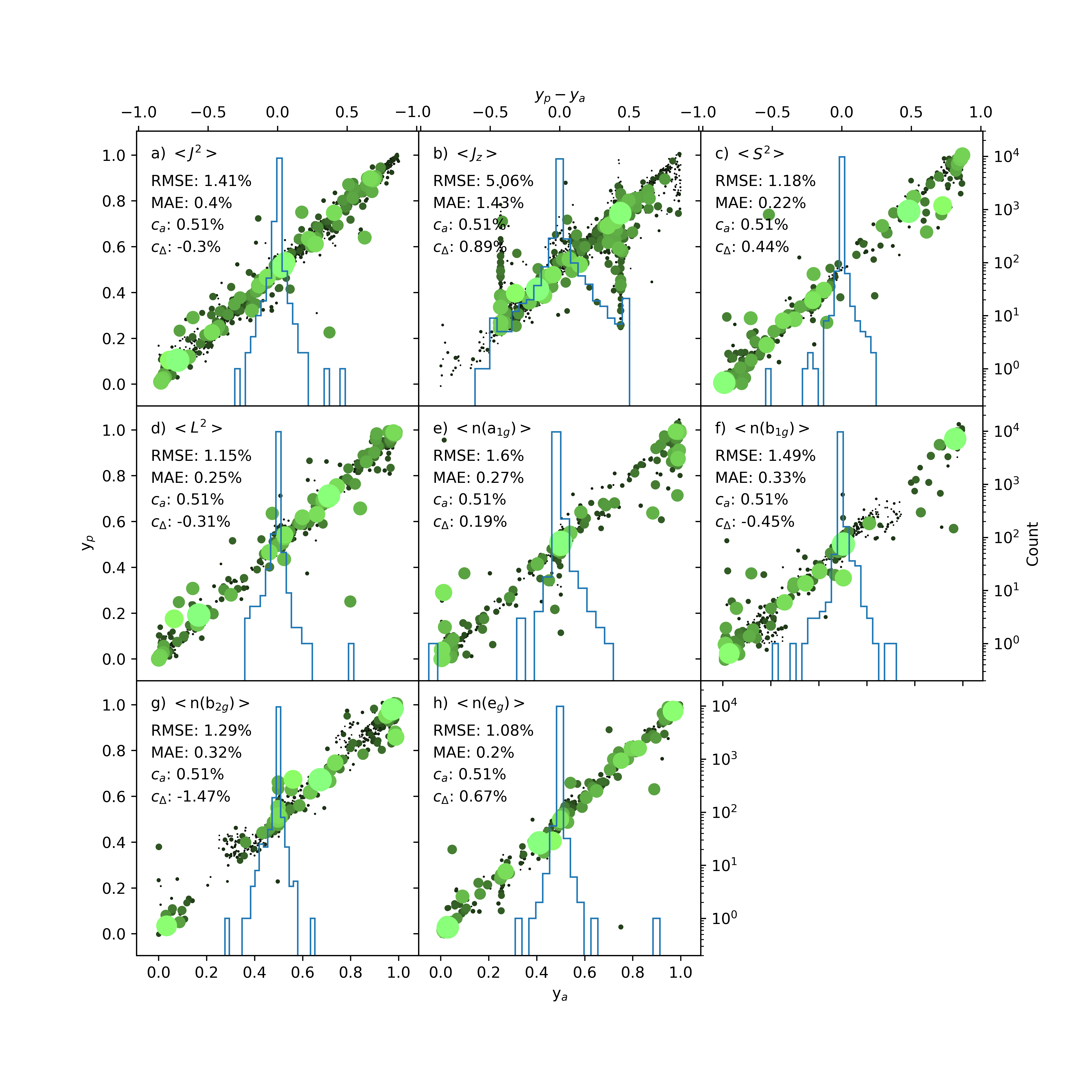}
\caption{(Color online) Comparison of normalized $y_p$ vs $y_a$ for predicted properties by A2pX-T2EP for Mn$^{2+}$ determined on test dataset (10000 spectra). The overlayed histograms show the distribution of  difference between $y_p$ vs $y_a$ on a logarithmic count scale. Pearson correlation to SE for $\Delta_y =y_p - y_a$ (c$_{\Delta}$) and y$_a$ (c$_a$). Color code and data point sizes are as in Fig.~\ref{fig:mn_ael}. The scales of the histograms are on the right  top, the scales of predicted (y$_p$) vs actual data (y$_a$) are given on the bottom  left. }
\label{fig:mn_pel}
\end{figure}

Figure~\ref{fig:ni_ael} shows the actual (y$_a$) vs the analysed (y$_p$) values for  D$_q$, D$_s$, D$_t$, S$_1$, S$_2$ $\Gamma_1$ and $\Gamma_2$ for  Ni$^{2+}$ obtained from A2pX-T2EP, i.e. the results are based on the dataset including variable T and FWHM as additional features in the input vector. 
Likewise, Figures~\ref{fig:fe_ael} and \ref{fig:mn_ael} show the results for Fe$^{2+}$ and Mn$^{2+}$, respectively.
Moreover, A2pX-T2EP simultaneously predicts expectation values of  $J^2$, $J_z$, $L^2$, $S^2$ and the occupations of d-states that are shown in Figures~\ref{fig:ni_pel} for Ni$^{2+}$, \ref{fig:fe_pel} for Fe$^{2+}$ and \ref{fig:mn_pel} for Mn$^{2+}$. The corresponding figures for Co$^{2+}$ are given in the main article. Tabel~\ref{tab:res02} in the main article summarizes  the accuracy of A2pX-T2EP for these elements.

\subsection{Prediction of Electronic States}

\subsection{Comparison between Reference Spectra and Recomputed Spectra}

\begin{figure}[!htbp]
\centering
\includegraphics[width=0.99\textwidth]{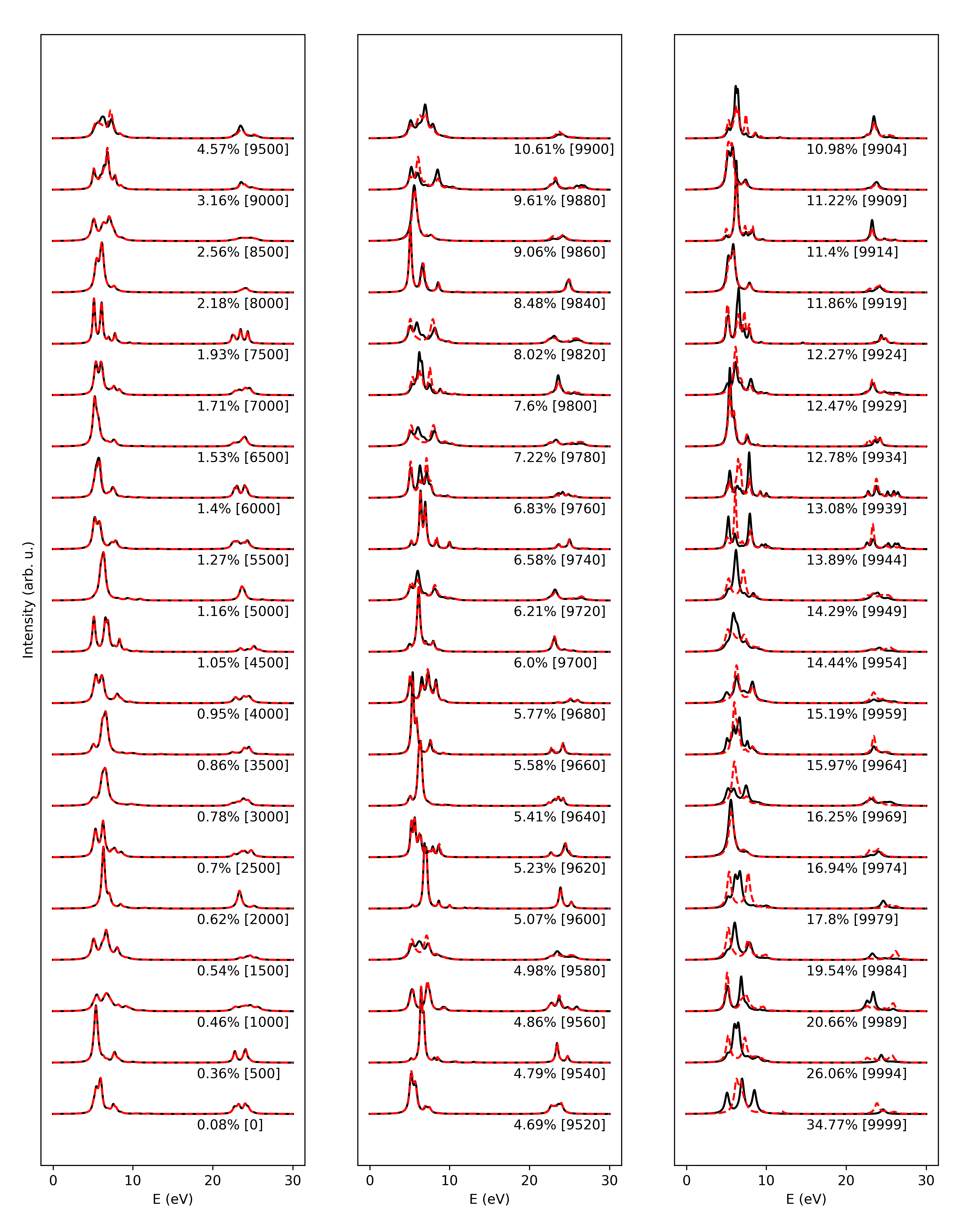}
\caption{ (Color online)  Overview of selected Ni$^{2+}$  reference spectra (black curve) and recomputed spectra (red dashed). The SE (\%) is indicated as well as the ranked position according to SE in square brackets in the test dataset. The energy scale is shifted.}
\label{fig:ni_comp}
\end{figure}

\begin{figure}[!htbp]
\centering
\includegraphics[width=0.99\textwidth]{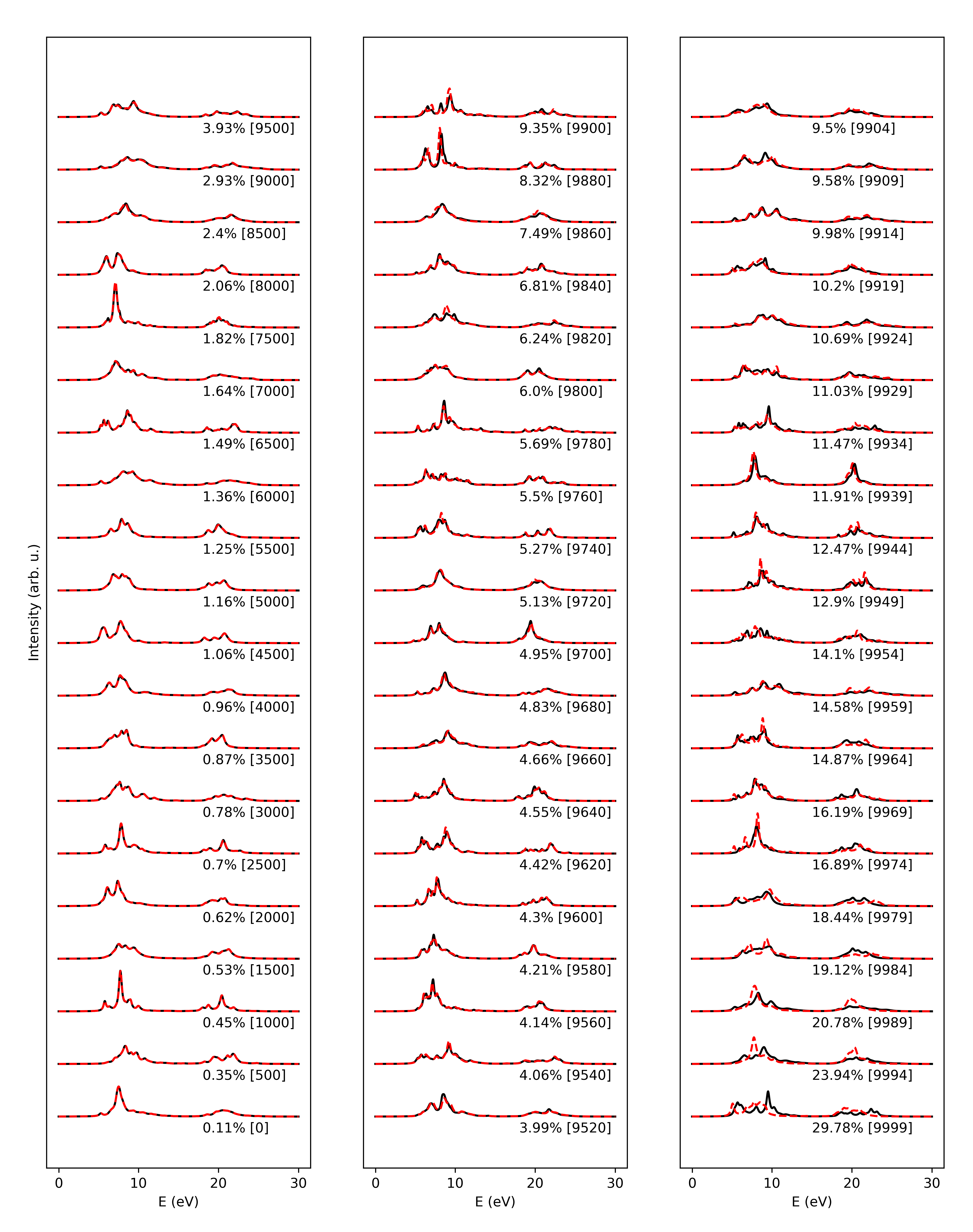}
\caption{(Color online) Overview of selected Fe$^{2+}$  reference spectra (black curve) and recomputed spectra (red dashed). The SE (\%) is indicated as well as the ranked position according to SE in square brackets in the test dataset. The energy scale is shifted.}
\label{fig:fe_comp}
\end{figure}

\begin{figure}[!htbp]
\centering
\includegraphics[width=0.99\textwidth]{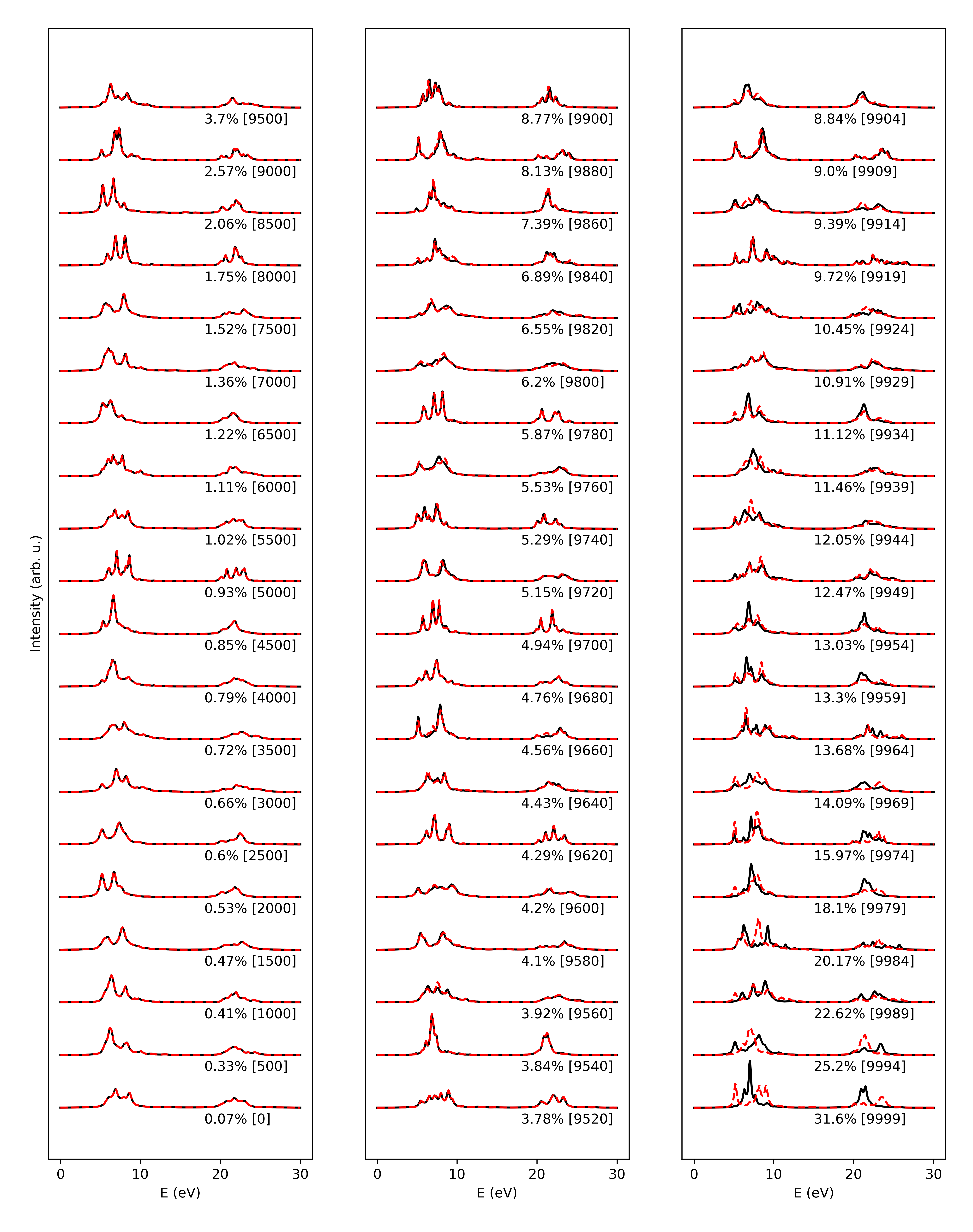}
\caption{(Color online)   Overview of selected Co$^{2+}$  reference spectra (black curve) and recomputed spectra (red dashed). The SE (\%) is indicated as well as the ranked position according to SE in square brackets in the test dataset. The energy scale is shifted.}
\label{fig:co_comp}
\end{figure}

\begin{figure}[!htbp]
\centering
\includegraphics[width=0.99\textwidth]{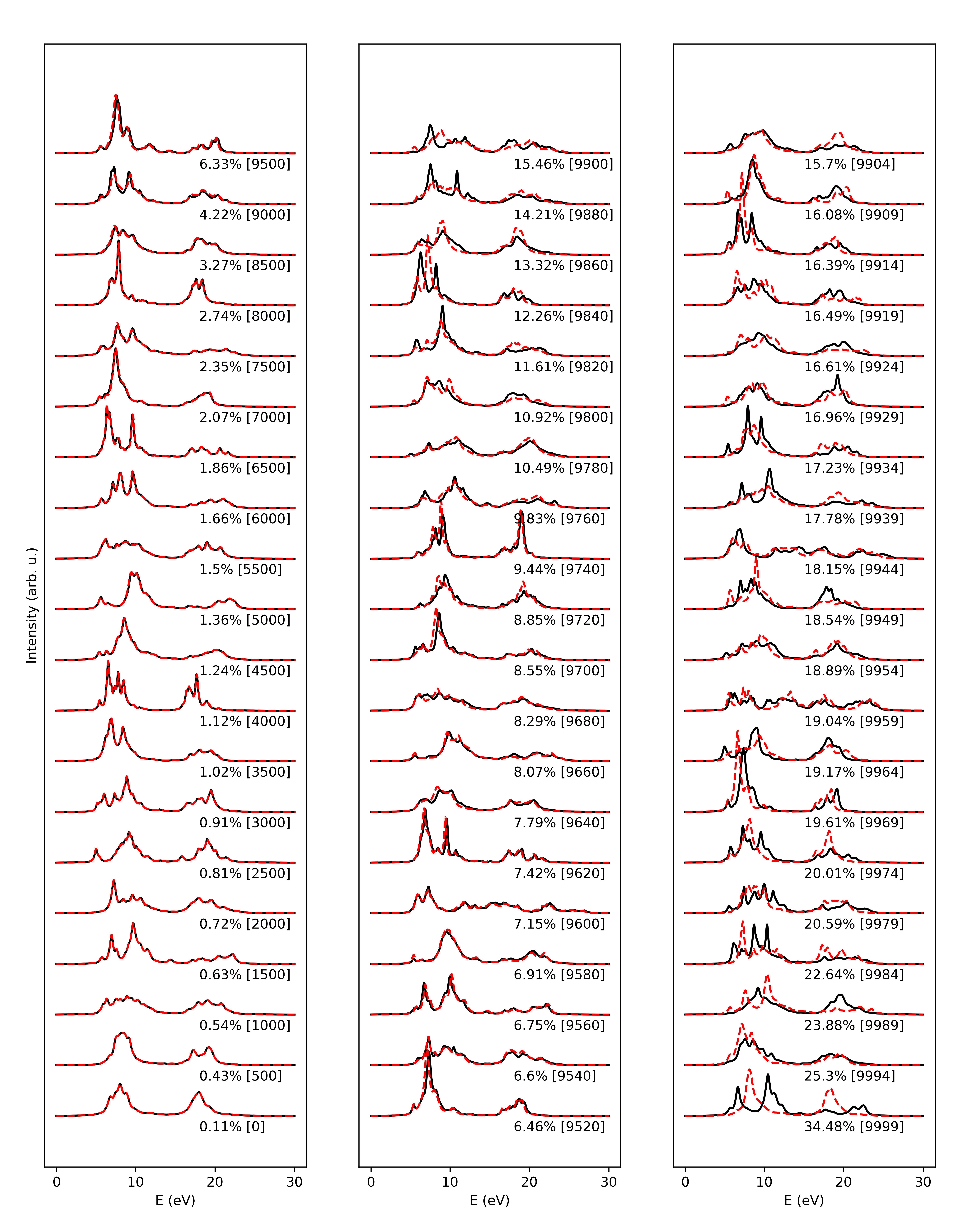}
\caption{(Color online)  Overview of selected Mn$^{2+}$  reference spectra (black curve) and recomputed spectra (red dashed). The SE (\%) is indicated as well as the ranked position according to SE in square brackets in the test dataset. The energy scale is shifted.}
\label{fig:mn_comp}
\end{figure}

Figures~\ref{fig:ni_comp} to \ref{fig:mn_comp} show  comparisons between reference spectra  (solid black) used to test the performance of the A2pX-T2EP and the corresponding recomputed spectra (red dashed) for Ni$^{2+}$,  Co$^{2+}$, Fe$^{2+}$ and Mn$^{2+}$, respectively. First, the ANNs  determined the parameters from the reference spectra. Then, the parameters were used in Eq.~(\ref{eq:ModelHAM}) and (\ref{eq:spect}) to yield the recomputed spectrum. This was done for all spectra in the test datasets. The spectrum errors (SE - defined in the main text) were computed and a selection of them spanning the whole range of SEs is given in the figures. SE (\%) and the ranking position from best (\textit{[0]}) to worst (\textit{[9999]})  are indicated in the as well.

\subsection{Influence of Noise}
\begin{figure}[!htbp]
\centering
\includegraphics[width=0.95\textwidth]{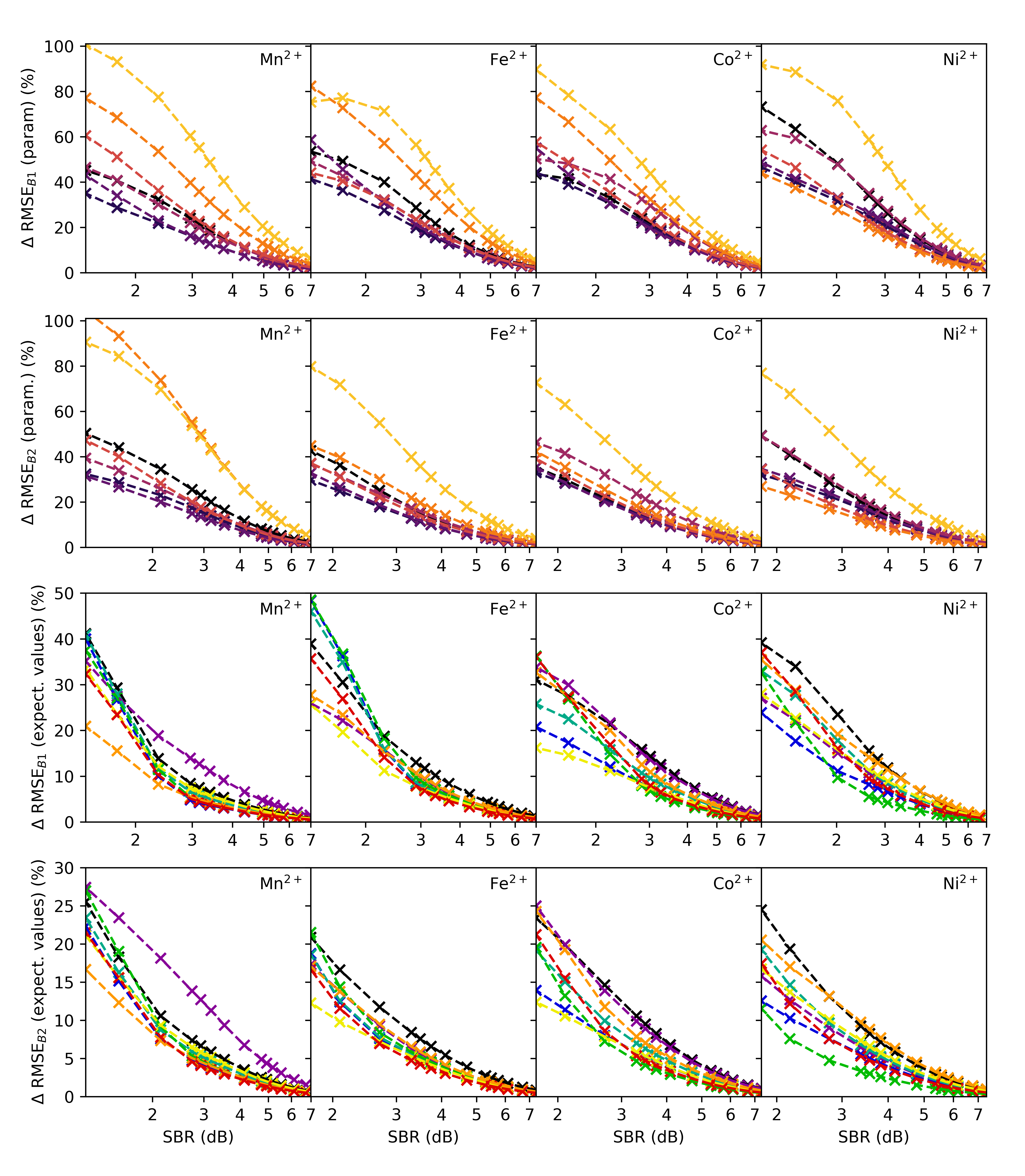}
\caption{(Color online) Influence of step background centered at either L$_3$ (B$_1$) or L$_2$ (B$_2$) in test spectra on the A2pX-T2EP's analytic (parameter estimation) and predictive (expectation value estimation) performance measured in $\Delta$RMSE. Same curve labeling as in Fig.~\ref{fig:rmse_noise} applies. }
\label{fig:stepbackground}
\end{figure}

Figure \ref{fig:stepbackground} shows the impact of different step backgrounds on the ANN results. It displays the impact of a step background on the $\Delta$RMSE in parameter and expectation value estimation. 

\end{document}